%% file: paper.tex
\documentclass[letterpaper,twocolumn,10pt]{article}
\usepackage{usenix2019_v3}
\usepackage{tikz}
\usepackage{amsmath}

\usepackage{filecontents}
\usepackage[linesnumbered,ruled,vlined]{algorithm2e}
\usepackage{algpseudocode}
\usepackage[linesnumbered]{algorithm2e}
\SetAlFnt{\small}
\algnewcommand\algorithmicforeach{\textbf{for each}}
\algdef{S}[FOR]{ForEach}[1]{\algorithmicforeach\ #1\ \algorithmicdo}

\usepackage[toc,page]{appendix}
\usepackage[export]{adjustbox}
\usepackage[normalem]{ulem}
\usepackage{graphicx}
\usepackage{multirow}
\usepackage{makecell}
\usepackage{booktabs}
\usepackage{subfig}
\usepackage{float}
\newcommand{\ra}[1]{\renewcommand{\arraystretch}{#1}}

\usepackage{tikz}

\usepackage{color}

\newcommand{\system}{Daydream\xspace}
\newcommand{\task}{task\xspace}
\newcommand{\uoft}{{\large$^\dag$}}
\newcommand{\msr}{{\large$^\star$}}

\usepackage[absolute]{textpos}
\newcommand{\ToAppear}{%
\begin{textblock*}{\textwidth}(0.95in,0.4in)
\end{textblock*}
}

\begin{document}

\ToAppear


\title{\system: Accurately Estimating the Efficacy of Optimizations for DNN Training
}
\author{Hongyu Zhu\uoft, Amar Phanishayee\msr, Gennady Pekhimenko\uoft \\
\rm{\textit{\uoft University of Toronto \& Vector Institute\hspace{0.02in}} \msr Microsoft Research\hspace{0.02in}} \\
}

\date{}
\maketitle

\thispagestyle{empty}

\input{sections/abstract.tex}
\input{sections/introduction.tex}
\input{sections/background.tex}

\input{sections/key_idea.tex}
\input{sections/design.tex}

\input{sections/methodology.tex}
\input{sections/evaluation.tex}
\input{sections/discussion}

\input{sections/related.tex}
\input{sections/conclusion.tex}
\input{sections/acknowledgement.tex}

\bibliographystyle{plain}
\bibliography{paper}

\input{sections/appendix.tex}

\end{document}

%% file: sections/abstract.tex
\begin{abstract}

Modern deep neural network (DNN) training jobs use complex and heterogeneous software/hardware stacks. The efficacy of software-level optimizations can vary significantly when used in different deployment configurations.
It is onerous and error-prone for ML practitioners and system developers to implement each optimization separately, and determine which ones will improve performance in their own configurations. 
Unfortunately, existing profiling tools do not aim to answer predictive questions such as "How will optimization X affect the performance of my model?".
We address this critical limitation, and proposes a new profiling tool, \system{}, to help programmers efficiently explore the efficacy of DNN optimizations. \system{} models DNN execution with a fine-grained dependency graph based on low-level traces collected by CUPTI~\cite{cupti}, and predicts runtime by simulating execution based on the dependency graph. \system{} maps the low-level traces using DNN domain-specific knowledge, and introduces a set of graph-transformation primitives that can easily model a wide variety of optimizations.
We show that \system{} is able to model most mainstream DNN optimization techniques, and accurately predict the efficacy of optimizations that will result in significant performance improvements.
\end{abstract}

%% file: sections/introduction.tex
\vspace{-2mm}
\section{Introduction}
\vspace{-2mm}

Recent years have witnessed the co-evolution of deep neural network (DNN) algorithms and the underlying hardware and software design.
ML researchers have developed many important models~\cite{devlin2018bert,he2017mask,resnet,vaswani2017attention} at a rapid pace, creating a huge demand for computation power~\cite{schwartz2019green}.
To meet the demand for fast DNN computation, computer architects respond with new, AI-optimized GPUs (e.g., NVidia Turing architecture~\cite{gpu-turing}) and various domain-specific hardware accelerators from FPGAs (e.g., Microsoft Catapult~\cite{putnam2014reconfigurable}) to ASICs (e.g., Google TPU~\cite{tpu}, Amazon Inferentia~\cite{inferentia}).
However these accelerators might not be effective in improving performance without proper software optimizations across the full systems stack~\cite{zhu2018benchmarking}.
As a result, systems researchers have proposed many optimizations, targeting different bottlenecks across the system stack \--- for example, improving memory utilization~\cite{vdnn,gist}, better overlapping of communication with computation~\cite{zhang2017poseidon,jayarajan2019priority,hashemi2018tictac}, and increasing communication efficiency~\cite{cho2019blueconnect}.
Moreover, researchers have also developed workload-centric optimizations to exploit the stochastic nature of DNN computation. 
For example, precision reduction~\cite{das2018mixed,gupta2015deep,micikevicius2017mixed} aims to reduce runtime as well as memory consumption, and gradient compression~\cite{lu2018multi,lin2017deep} aims at reducing the communication overhead in distributed training. 

Despite these advances, the benefits of many proposed optimizations cannot be fully exploited due to two main reasons.
First, the efficacy of many proposed performance optimizations can drastically change when applied to different ML models and deployment configurations. 
The hardware deployments that practitioners use might be completely different from the hardware configurations used by optimization and model inventors.
Differences in DNN models, accelerator type, compute capabilities, available memory, networking capabilities, and software library versions can all shift the major runtime bottlenecks.
Second, it is onerous for programmers to implement and evaluate various optimizations to identify the ones that actually work for their models.
As a result, it is common for users to ask \emph{what-if} questions such as:

\textit{Why did my DNN training workload run slowly?  Will optimization X improve the performance of my model?
Does GPU memory capacity limit the performance of my model?  
Would upgrading to a faster network improve training throughput?
How will my workload scale with the number of GPUs? 
}

The central focus of this paper is to answer the following general question for DNN training workloads: \textit{Given a model and a deployment scenario, how can we efficiently explore the efficacy of potential solutions}?
Systems researchers have tried to explore the impact of different potential performance bottlenecks (e.g., CPU, network, IO) in many non-ML contexts~\cite{miller1987ips,aguilera2003performance,von2008modeling,curtsinger2015c,ousterhout2015making,ousterhout2017monotasks}. The basic approaches to explore the what-if questions are similar: decompose the workloads into atomic tasks, profile runtime statistics for each task, model the what-if question, and use simulation to estimate performance.
These systems typically address what-if questions of the form: "How does runtime change if a task $T$ is $N$ times (or even infinitely) faster?"~\cite{curtsinger2015c,ousterhout2015making}.  Such questions can be simply modeled by shrinking \task{} runtime.
While this basic approach seems sufficient to address the central question above for ML workloads,
the \textbf{diversity of DNN optimizations} introduces three key requirements unique to these workloads,
thus motivating the need for a novel solution.

First, we need to \textbf{track dependencies at a kernel-level abstraction} 
i.e., one GPU kernel corresponds to one \task{} (the smallest unit of execution in the dependency graph).
Such fine-grained abstraction is necessary because optimizations that improve hardware utilization typically target individual compute kernels (e.g., mixed precision~\cite{micikevicius2017mixed}).
Meanwhile, accurate performance estimation has to consider both CPU and GPU runtime. Certain optimizations, e.g., kernel fusion, require potentially removing existing CPU and GPU tasks from the dependency graph.
Existing tools do not provide such dependency tracking.
It is therefore important to track kernel-level dependencies among concurrently executing \task{}s.


Second, we need to \textbf{map \task{}s to DNN layers}. In contrast to prior works that explore what-if questions in non-ML contexts, predicting the performance of DNN optimizations requires domain knowledge about DNNs  to properly model them.
For example, MetaFlow~\cite{jia2019optimizing} and TASO~\cite{jia2019taso} fuse DNN layers. Modeling them requires a mapping from \task{}s to specific DNN layers.
However, collecting kernel-level traces on accelerators requires generic vendor-provided tools (e.g., NVProf~\cite{nvprof}, CUPTI~\cite{cupti}), which have no application specific knowledge.
We therefore need to have the ability to map low-level \task{}s to DNN layers.

Third, we need the \textbf{ability to easily model diverse DNN optimizations}. 
Modeling a DNN optimization might involve not just scaling or shrinking task durations, but also complicated transformations to the dependency graph. For example, TicTac~\cite{hashemi2018tictac} reschedules communication tasks, BlueConnect~\cite{cho2019blueconnect} replaces the communication primitives to utilize parallel network channels, and the optimization proposed by Jung~\emph{et al.}~\cite{jung2018restructuring} restructures the GPU kernel implementations. 
Manually manipulating the kernel-level dependency graph could be extremely intricate and error-prone.
The system should enable users to flexibly and effectively model such diverse optimizations with minimal effort.

We introduce \emph{\system{}}, a new system that fulfills all three requirements described above, and achieves our goal of answering potential what-if questions for DNN workloads.
Constructing dependencies among potentially thousands of low-level \task{}s is not an easy problem: tasks can be spread across multiple execution threads (including both CPU threads and GPU streams), thus even for simple DNN workloads, this results in thousands of tasks to be tracked.
The intricacy comes from identifying dependencies across threads. We make a key observation about DNN training workloads: despite the large number of tasks that need to be tracked, the number of concurrently executing threads is surprisingly quite limited.
Based on this observation, \system{} constructs the low-level dependency graph, which provides a realistic model of overlapping among CPU, GPU, and communication runtimes in a DNN training workload. 
It uses a synchronization-free approach to map GPU tasks onto appropriate higher-level DNN layer abstractions. 
We also introduce a set of graph-transformation rules, allowing programmers to effectively model various performance optimizations.
After modeling the optimization, \system{} simulates the execution based on the new dependency graph to predict the overall runtime.
In our evaluation, we show that \system{} is able to distinguish effective DNN optimizations from those that will bring limited improvements by accurately predicting their performance speedups.

In summary, we make the following key contributions:

\begin{itemize}
\vspace{-2mm}
    \item We make the observation that fine-grained \task{}s in DNN training workloads are highly sequential.
    This greatly simplifies dependency graph construction, over thousands of tasks, as we only need to identify a limited number of inter-thread dependencies.
\vspace{-2mm}
    \item Daydream introduces the abstraction of a kernel-granularity dependency graph that contains mappings back to DNN specific abstractions (layers), by collecting profiling data, instrumenting DNN frameworks, and exploiting information from vendor-provided tools like CUPTI.  
Daydream also provides primitives to mutate the dependency graph in the form of simple graph transformations.  
Taken together this enables programmers to both (i) model a diverse set of popular optimizations spanning kernel- and layer-level enhancements by using simple graph-transformation primitives, and (ii) estimate the efficacy of optimizations by simulating execution time based on optimization-induced graph mutations.
\vspace{-2mm}
    \item We extensively evaluate \system{}, with \emph{five} different optimizations on \emph{five} DNN models across \emph{three} distinct applications. We show that \system{} can effectively detect which optimizations provide improvements and also accurately predict their magnitude for different DNN models and deployments.
    For example, we estimate that using mixed precision will improve the iteration time of training BERT\textsubscript{LARGE} model by 17.2\% (with <3\% error), while the kernel fusion technique can improve it by 38.7\% (with <7\% error).
    We can also accurately predict performance in distributed training with different number of workers and variable network bandwidth, based on runtime profiles collected from a single-GPU setting.
\end{itemize}

%% file: sections/background.tex
\vspace{-2mm}
\section{Background}

\begin{table*}
\ra{1.1}
\small
\centering
\begin{tabular}{lll}
\toprule
\textbf{Optimization Goal}                                                                          & \textbf{Strategy}                                                                               & \textbf{Technique Examples}                                                                                                                           \\ \hline
\multirow{5}{*}{\begin{tabular}[c]{@{}l@{}}Improving Hardware Utilization\\ in Single-Worker Setting\end{tabular}}  & \begin{tabular}[c]{@{}l@{}}Increasing Mini-batch Size by\\Reducing Memory Footprints\end{tabular}                                                             & \textbf{vDNN}~\cite{vdnn}, \textbf{Gist}~\cite{gist}, Chen \emph{et al.}~\cite{chen2016training}                                                                                                                      \\ \cline{2-3} 
                                                                                           & Reducing Precision                                                                       & \textcolor{blue}{\textbf{\textit{Micikevicius} \emph{et al.}}}~\cite{micikevicius2017mixed}, Gupta~\emph{et al.}~\cite{gupta2015deep}, Das~\emph{et al.}~\cite{das2018mixed}                                                                                                                      \\ \cline{2-3}
                                                                                           & Fusing Kernels/Layers                                                                         & \textcolor{blue}{\textbf{\textit{FusedAdam}}}~\cite{apex_opt}, \textbf{MetaFlow}~\cite{jia2019optimizing}, Ashari \emph{et al.}~\cite{ashari2015optimizing}, TASO~\cite{jia2019taso}                                                                                                                      \\ \cline{2-3}
                                                                                           & \begin{tabular}[c]{@{}l@{}}Improving Low-level Kernel\\ Implementation\end{tabular}  &
                                  \begin{tabular}[c]{@{}l@{}} \textcolor{blue}{\textit{\textbf{Restructing Batchnorm}}}~\cite{jung2018restructuring}, Tensor Comprehensions~\cite{vasilache2018tensor},\\Kjolstad \emph{et al.}~\cite{kjolstad2017tensor}, TVM~\cite{chen2018tvm} \end{tabular}                                                           \\ \hline
\multirow{3}{*}{\begin{tabular}[c]{@{}l@{}}Lowering Communication Overhead\\ in Distributed Training\end{tabular}} & \begin{tabular}[c]{@{}l@{}}Reducing Communication\\ Workloads\end{tabular}             & \begin{tabular}[c]{@{}l@{}}\textbf{Deep Gradient Compression}~\cite{lin2017deep}, AdaComm~\cite{wang2018adaptive}, Parallax~\cite{kim2019parallax},\\TernGrad~\cite{wen2017terngrad}, QSGD~\cite{alistarh2017qsgd}\end{tabular} \\ \cline{2-3} 
                                                                                           & \begin{tabular}[c]{@{}l@{}}Improving Communication\\ Efficiency/Overlap\end{tabular} & \begin{tabular}[c]{@{}l@{}}\textcolor{blue}{\textit{\textbf{Wait-free Backprop}}}~\cite{zhang2017poseidon}, \textcolor{blue}{\textit{\textbf{P3}}}~\cite{jayarajan2019priority}, \textbf{BlueConnect}~\cite{cho2019blueconnect}, TicTac~\cite{hashemi2018tictac},\\     BytePS~\cite{peng2019generic}, Xue~\emph{et al.}~\cite{xue2019fast} \end{tabular}\\ 
\bottomrule
\end{tabular}
\caption{Representative optimizations for DNN training. We show how we can accurately estimate the performance of optimizations (shown in \textcolor{blue}{\textbf{\textit{italics}}}) in Section~\ref{sec:eval}, and can effectively model many other optimizations (shown in \textbf{bold}) in Section~\ref{sec:reduce}.}
\vspace{-3mm}
\label{table:approches}
\end{table*}

\vspace{-1mm}
\subsection{DNN Training Basics}
\vspace{-1mm}

DNN training is an iterative algorithm, in which one iteration consists of three phases: (i)~\emph{forward}, (ii)~\emph{backward}, and (iii)~\emph{weight update}. 
The \emph{forward} phase takes training data samples as input and produces output based on current weights (or parameters). 
The error between the \emph{forward} output and the input data labels is fed to the \emph{backward} phase, which computes the gradients of weights with respect to the input data. 
The \emph{weight update} phase then uses the gradients to update weights accordingly. 
In each iteration, the input data samples are randomly selected~\cite{bottou2010large}, forming a~\emph{mini-batch} of input.


\vspace{-2mm}
\subsection{DNN Training Optimizations}
\vspace{-1mm}

Modern DNNs have millions of parameters~\cite{deep_compression}, resulting in training times of days or even weeks~\cite{krizhevsky2012imagenet}. 
To improve DNN training performance, researchers have proposed various strategies focusing on different optimization goals. 
To understand the potential what-if questions and how to design a system to answer them, we study a list of software-level techniques that speedup DNN training from top systems and ML conferences in recent years. Table~\ref{table:approches} shows our summary.

\textbf{Exploiting computation power of hardware accelerators.} ML programmers often use large mini-batches, within the memory budget, for better hardware utilization and faster convergence.
This motivates strategies that reduce the memory footprint of DNN training and hence enables training with larger mini-batch sizes~\cite{chen2016training,gist,vdnn}. Researchers have also proposed some generic strategies to increase hardware utilization, including precision reduction~\cite{das2018mixed,gupta2015deep,micikevicius2017mixed}, kernel/layer fusion~\cite{ashari2015optimizing,jia2019optimizing,jia2019taso}, and improving low-level kernel implementation~\cite{chen2018tvm,jung2018restructuring,kjolstad2017tensor,vasilache2018tensor}. Meanwhile, libraries such as cuDNN~\cite{chetlur2014cudnn}, cuBLAS~\cite{cublas}, MKL~\cite{wang2014intel}, Eigen~\cite{eigen}, NCCL~\cite{nccl}, are also constantly evolving to provide operations and primitives that can better utilize underlying hardware.

\textbf{Scalable distributed training.} Data parallelism~\cite{bottou2010large} is a simple and effective strategy to improve training performance. Using multiple accelerators significantly reduces DNN training time to hours or even minutes~\cite{mlperf-training-0.6}. 
This success is mainly based on the techniques that guarantee model convergence under extremely large mini-batch size~\cite{akiba2017extremely,goyal2017accurate,you2018imagenet}.
One of the major performance bottlenecks for distributed training is communication, which can be optimized by compressing traffic~\cite{lin2017deep,lu2018multi,wang2018adaptive,wen2017terngrad}, increasing network utilization~\cite{cho2019blueconnect,xue2019fast}, or increasing the overlap between communication and computation~\cite{hashemi2018tictac,jayarajan2019priority,zhang2017poseidon}. 
Exploring the efficacy of these optimizations without prediction requires a multi-machine cluster. Our proposed design, \system{}, avoids the potential cost of cluster setup (i.e. extra machines, accelerators, high-speed communication), by predicting distributed training performance with profiles collected from a single-worker environment.


\vspace{-2mm}
\subsection{Profiling Tools for DNNs}
\vspace{-1mm}

As the full ML system stack is constantly evolving, profiling tools play a key role in helping programmers identify the performance bottlenecks under different system configurations.

\textbf{Hardware profiling tools.} Modern DNN training heavily relies on hardware accelerators such as GPUs~\cite{gpu-turing} and TPUs~\cite{tpu}. To help programmers develop highly efficient applications, hardware vendors provide profiling tools that can expose hardware performance counters. For example, NVProf~\cite{nvprof} provides programmers with information including start/end time, core utilization, memory throughput, cache miss rate, along with hundreds of other hardware counters for every GPU kernel. CUPTI~\cite{cupti} enables programmers to extract and manipulate these counters at runtime. Nsight~\cite{nsight} aims to provide details on the state of more fine-grained counters for recent GPU architectures~\cite{gpu-turing}.
Our proposed system, \system{}, relies on CUPTI to collect low-level traces for further analysis. 


\textbf{Framework built-in tools.} For more intuitive profiling results, it is often desirable for a profiler to show runtime statistics for framework operations, or even DNN layers. DNN frameworks have built-in tools to achieve this goal by correlating the hardware counters with runtime information collected in frameworks. TensorFlow~\cite{tf}, coupled with the Cloud TPU Tool~\cite{tputool}, can provide an execution timeline and runtime statistics for each TensorFlow operation. Similarly, other mainstream frameworks (e.g., MXNet~\cite{mxnet} and PyTorch~\cite{pytorch}) provide built-in tools that can extract per-layer or per-operation runtime from both the CPU and the GPU. The framework built-in tools render intuitive results for programmers, but omit important details (for example, the CPU runtime). We show in our work that such information is crucial in building an accurate runtime predictor.



%% file: sections/key_idea.tex
\vspace{-2mm}
\section{Key Ideas} \label{sec:idea}
\vspace{-1mm}

In this section we highlight the key ideas and observations behind the \system{} design.

\textbf{Constructing kernel-granularity dependency graph.} The neural network topology is a natural graph structure in which nodes are DNN operators or layers. Most mainstream DNN frameworks~\cite{mxnet,pytorch}
provide built-in tools to record the layer-level runtime profile.
The layer-level abstraction is intuitive for programmers to understand the "where time goes" question, but hides important information about the parallel execution of the CPU functions, GPU kernels, and memory transfers.
This information is crucial for accurate performance predictions. For example, optimizations that reduce numerical precision will change the duration of GPU kernels while the CPU runtime remains unchanged, and optimizations like vDNN~\cite{vdnn} will inject CUDA memory copies, without changing the duration of GPU kernels.
It is extremely hard to predict how duration of each layer changes when applying these optimizations if lacking low-level details about CPU and GPU runtime.
To accommodate optimizations that target fine granularity tasks (such as GPU kernels), our proposed system, \system{} chooses to model the training workloads using a kernel-level dependency graph (i.e., each GPU kernel has one corresponding \task{} in the graph), incorporating detailed traces of CPU, GPU and communication runtime.

\begin{figure}
    \includegraphics[width=\columnwidth]{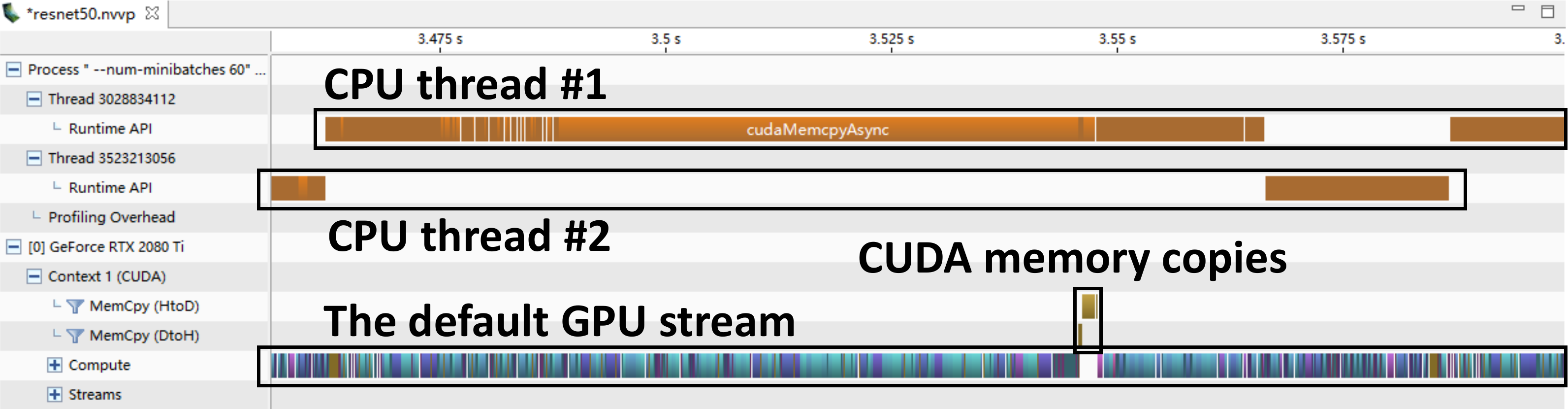}
    \caption{NVProf timeline example of training ResNet-50.}
    \label{fig:nvprof_screenshot}
    \vspace{-4mm}
\end{figure}

With a large number of kernel-level \task{}s that are spread across several threads and CUDA streams, the complexity of constructing the dependency graph comes mainly from identifying the inter-thread dependencies~\cite{von2008modeling}.
Existing tools do not provide such dependency tracking.
We make the following key observations about the DNN training workloads to overcome this general challenge of dependency tracking in concurrent systems.
First, for the implementations in the mainstream frameworks~\cite{mxnet,pytorch}, once a mini-batch has been prepared by data loading threads, only one or two CPU threads are involved in the control flow of computation.\footnote{Our approach can be generalized to frameworks that use more concurrent CPU threads.}
Second, there is a very limited number of concurrent GPU kernels. Such serialization of GPU kernels is due to two main reasons: (i) GPU kernels in the modern cuDNN library achieve high GPU core utilization; (ii) ML frameworks usually invoke only one CUDA stream.
Figure~\ref{fig:nvprof_screenshot} shows the NVProf profiles of one training iteration of ResNet-50.
There are two CPU threads involved, but no CPU \task{}s run concurrently. The high serialization of low-level traces is not a unique phenomenon for just convolutional networks. We observe a similar phenomenon in most DNN training workloads.

Based on these insights, \system{} constructs the kernel-level dependency graph in three major steps. First, \system{} uses CUPTI to extract traces of all GPU kernels, CUDA memory copies, and CUDA APIs. Second, \system{} captures the dependencies between CPU and GPU \task{}s, caused by CUDA synchronizations and GPU kernel launches. Third, when predicting performance for distributed training, \system{} adds communication \task{}s to the dependency graph.


\textbf{Synchronization-free \task{}-to-layer mapping.} In distributed training, mainstream frameworks implement the wait-free backpropagation strategy~\cite{zhang2017poseidon} to overlap communication with computation.
This strategy immediately transfers gradients once they are computed by corresponding backward layers. 
To properly add dependencies related to communication \task{}s, we need the \task{}-to-layer mapping to know when the computation of each layer ends.
Meanwhile, accurately modeling DNN optimizations by changing the graph potentially requires this \task{}-to-layer mapping to determine which \task{}s are involved and how to change them.

Unfortunately, vendor-provided tools like CUPTI do not have the required knowledge about these applications and building such a mapping requires extra DNN framework instrumentation.
A na\"ive approach to achieve this mapping is to compare the start and stop timestamps of GPU kernels and DNN layers.
This requires additional CUDA synchronization calls for each layer since GPU kernels are launched asynchronously. However, such synchronizations might significantly alter the execution runtime by adding additional dependencies from GPU to CPU \task{}s. Hence, we design a synchronization-free procedure to achieve this mapping by instrumenting timestamps for each layer in the frameworks, and utilizing the correlations between CPU and GPU \task{}s. 

\textbf{Representing complex optimizations with simple graph-transformation primitives.} As shown in Table~\ref{table:approches}, DNN optimizations target a wide range of performance bottlenecks with various approaches.
Unlike prior dependency graph analysis in non-ML contexts~\cite{curtsinger2015c,ousterhout2015making,ousterhout2017monotasks}, where users can model most what-if questions by simply shrinking and scaling \task{} runtime, accurately modeling DNN optimizations with the low-level dependency graph might require complicated changes to the dependency graph.
Manually changing the kernel-level graph to model optimizations could be both complicated and error-prone, and the programmers might simply opt to rather directly implement the optimizations.

To address this problem, we propose a small set of graph-transformation primitives, so that popular optimization techniques can be effectively represented as a combination of these primitives.
These primitives include (i) \task{} insertion/removal, (ii) \task selection and update, and (iii) changing the policy for scheduling \task{}s.
The proposed primitives are simple yet powerful enough to represent many different optimizations as we will show in Section~\ref{sec:reduce}. They play a key role in realizing our goal of efficiently exploring what-if questions.

In summary, Daydream introduces the abstraction of a kernel-granularity dependency graph that contains mappings back to DNN specific abstractions (layers).
It tracks dependencies by collecting profiling data as well as instrumenting DNN frameworks.  
Daydream also provides primitives to mutate the dependency graph in the form of simple graph transformations.  
Altogether this enables programmers to both (i) model a diverse set of popular optimizations spanning kernel- and layer-level enhancements by using simple graph-transformation primitives, and (ii) estimate the efficacy of optimizations by simulating execution time based on optimization-induced graph mutations.


%% file: sections/design.tex
\vspace{-2mm}
\section{Design} \label{sec:impl}
\vspace{-1mm}

We describe \system{}'s design with an emphasis on how to construct \system{}'s proposed graph abstraction: the kernel-granularity dependency graph with mappings back to DNN layers. We also describe the primitives for mutating this graph to model different optimizations and how \system{} uses the graph to estimate the efficacy of various DNN optimizations.

\vspace{-2mm}
\subsection{Overview of \system{}} \label{sec:idea:overview}
\vspace{-1mm}
\begin{figure*}
    \subfloat[Constructing the dependency graph based on CUPTI traces (the black arrows represent \task{} dependencies). \label{fig:overview_a}]{\includegraphics[width=0.65\columnwidth,valign=t]{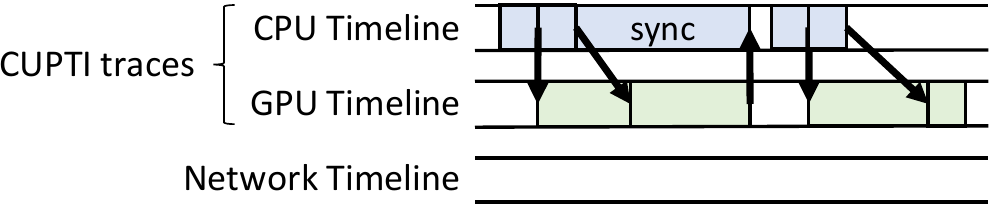}}
    \hspace{0.3cm}
    \subfloat[Mapping each \task{} to DNN layers (shown in different colors in the figure), then inserting communication tasks based on mapping and instrumentation. \label{fig:overview_a}]{\includegraphics[width=0.7\columnwidth,valign=t]{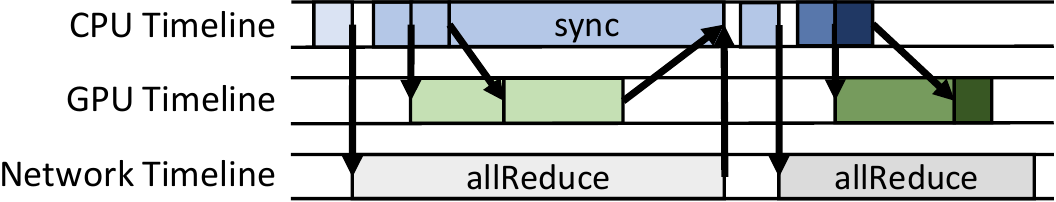}}
    \hspace{0.3cm}
    \subfloat[Predicting "what if network bandwidth is 2$\times$" by shrinking allReduce duration by 2$\times$ and simulating the new dependency graph.\label{fig:overview_c}]{\includegraphics[width=0.6\columnwidth,valign=t]{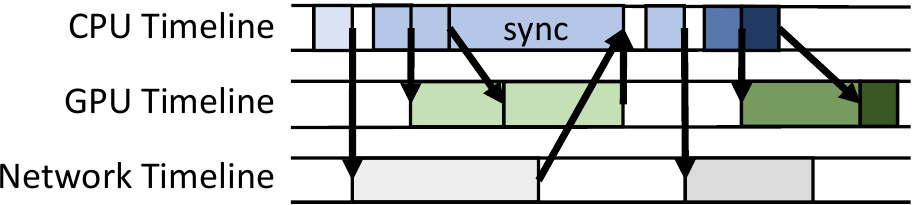}}
    \caption{An example showing \system{}'s overall workflow for predicting runtime assuming network bandwidth doubles.}
    \vspace{-4mm}
    \label{fig:overview}
\end{figure*}

Figure~\ref{fig:overview} shows the workflow of performance prediction in \system{}.  It consists of the following four phases:

\textbf{Phase 1: Trace collection.} Constructing a kernel-level dependency graph requires low-level details for all \task{}s. These details are extremely massive, differ across ML frameworks, and can be obtained by profiling a baseline workload.
\system{} collects low-level profiling data using CUPTI~\cite{cupti}, a tool which provides details for all CPU/GPU \task{}s including name, start time, duration, CUDA stream ID, thread ID, etc. We manually augment three popular frameworks (Caffe, MXNet, PyTorch) for use with CUPTI and modify the layer modules of these frameworks to collect timestamps of each layer, which will be used for \task{}-to-layer mapping, described in Section~\ref{sec:impl:map}.
Through our instrumentation, we also collect the necessary information (e.g., size of gradients) to construct the dependency graph of distributed training via a profile collected in a single worker setting.

\textbf{Phase 2: Dependency graph construction.} \system{} constructs the dependency graph with details of \task{}s provided by the first phase. A dependency could be induced by domain knowledge (e.g., a GPU \task{} triggers a communication \task{}), or by hardware/software implementation (e.g., a cudaLaunchKernel API triggers the corresponding GPU \task{}). Based on our analysis, we identify five different types of dependencies (described in Section~\ref{sec:impl:dep}), which are sufficient for \system{} to accurately simulate baseline execution. 

\textbf{Phase 3: Graph transformation.} To estimate the efficacy of a given optimization, \system{} models the optimization by transforming the dependency graph. \system{} provides a set of primitives to represent these transformations. We design these primitives in a way such that they are succinct (easy to use), flexible (able to depict a wide range of optimizations), and accurate (being able to achieve high prediction accuracy).

\IncMargin{1em}
\vspace{-2mm}
\begin{algorithm}
\DontPrintSemicolon
\SetAlgoLined
\SetKwInOut{Input}{Input}\SetKwInOut{Output}{Output}
\Indm
\Input{Dependency graph: $G(V, E)$}
\Output{The start time of each \task{} $u \in V$}
\BlankLine
\Indp

$F \gets \emptyset$ \tcp{initialize the frontier \task{} set}

$P \gets \{0\}$ \tcp{initialize thread progress}

\ForEach{\task{} $u \in V$}{
    $u.ref \gets |\{u's parents\}|$
    
    \uIf{$u$.ref = 0}{
        $F \gets F \cup \{u\}$
    }
}

\While{$F \neq \emptyset$}{
    $u \gets schedule(F)$ \tcp{pick a \task{} to exec.}
    
    $t \gets u.ExecutionThread$
    
    $F \gets F-\{u\}$
    
    $u.start \gets max(P[t], u.start)$
    
    $P[t] \gets u.start + u.duration + u.gap$ 
    
    \ForEach{$c \in u.children$}{
        $c.ref \gets c.ref - 1$

        $c.start \gets max(c.start, u.start + u.duration + u.gap)$

        \If{$c.ref=0$} {$F \gets F \cup \{c\}$}
    }
}
\caption{\system{}'s Simulation Algorithm}\label{algo:simulation}
\end{algorithm}
\vspace{-3mm}
\DecMargin{1em}

\textbf{Phase 4: Runtime simulation.} \system{} simulates the execution of optimizations to predict runtime based on the dependency graph. Algorithm~\ref{algo:simulation} shows the simulation process, which traverses the dependency graph and puts \task{}s into execution threads. In each iteration, \system{} picks one \task{} from the execution frontier (i.e. \task{}s that are ready to execute), dispatches it to its corresponding execution thread, and updates the thread progress. The simulation determines the start time of each \task{} and records the total execution time.


\vspace{-2mm}
\subsection{Dependency Graph Construction} \label{sec:impl:construct}
\vspace{-1mm}

Constructing the dependency graph is essentially to determine the node (\task{}) set and edge (dependency) set.

\vspace{-2mm}

\subsubsection{Task} \label{sec:impl:gran}
\vspace{-1mm}

\system{}'s kernel-level dependency graph contains the following four types of \task{}s:

\textbf{GPU \task{}s.} Each GPU \task{} in the graph corresponds to one GPU kernel. \system{} also views CUDA memory copies as GPU \task{}s, because each memory copy is associated with a specific CUDA stream, and therefore has dependencies with other GPU kernels. The runtime of all these tasks can be collected using CUPTI.

\textbf{CPU \task{}s.} To model the concurrency and dependencies between CPU runtime and the GPU runtime, \system{} generates CPU \task{}s based on CPU traces collected by CUPTI. One of the limitations of CUPTI is that it can only expose CUDA-related traces. Instead of adding massive instrumentation to the framework, \system{} captures the non-CUDA runtime by recording the lengths of gaps between consecutive CPU \task{}s (shown in line 13 of Algorithm~\ref{algo:simulation}).

\textbf{Data loading \task{}s.} One data loading \task{} corresponds to loading one mini-batch from disk/flash to CPU memory. We include data loading tasks for completeness, even though data loading in most DNN training workloads is not a performance bottleneck. In \system{}'s implementation, we treat all data loading \task{}s as CPU \task{}s. 

\textbf{Communication \task{}s.} A communication \task{} corresponds to one communication primitive, e.g., a push/pull operation in parameter-server based frameworks~\cite{li2014scaling}, or an all-reduce operation in decentralized frameworks.
When predicting distributed training performance, \system{} automatically adds communication tasks to the dependency graph based on a single-worker profile.
We notice that in PyTorch, gradients from multiple layers can be grouped and sent with a single allReduce primitive~\cite{pytorch-doc}.
Thus, properly adding communication tasks to a PyTorch profile requires additional instrumentation to extract knowledge about gradients grouping.
 
Given the types of \task{}s in the graph, \system{} collects and maintains the following information for each \task{}, which is later used in what-if analysis and simulation:


\textbf{ExecutionThread.} Depending on the type of a \task{}, its execution thread can be on of the following: (i) a CPU process, (ii) a GPU stream, and (iii) a communication channel. A data loading \task{} is executed in a CPU process. A CPU process has a process ID, a GPU stream has a stream ID, and a communication channel could be send/receive when using parameter server primitives, or a unified one when using collective primitives. This field is used in line 10 of Algorithm~\ref{algo:simulation}.

\textbf{Duration.} This field specifies how long a \task{} takes to execute. The duration of a CPU/GPU \task{} is collected by CUPTI. The runtime of data loading \task{}s is measured by injecting timestamps to the framework.
\system{} aims to predict distributed training performance based on profiling in a single-GPU configuration. Hence we calculate the duration of all communication \task{} based on the size of gradients, the communication type (push/pull/all-reduce), and the network bandwidth. These numbers can be obtained based on knowledge of the DNN model and framework implementation.

\textbf{Gap.} The duration of low-level CUDA APIs (e.g., \texttt{cudaMalloc}) might be only tens of microseconds, which is of the same magnitude as the runtime of their non-CUDA equivalent C functions (e.g., \texttt{malloc}), or the runtime of the call stack from Python front-end to C back-end. NVidia-provided tools cannot expose non-CUDA traces, but they are indispensable to simulation accuracy.
The non-CUDA CPU runtime is usually not a target for optimization in DNN models, hence, we do not need to define and measure corresponding \task{}s. Instead, for each CPU \task{} in our current definition, we measure the gap between its end and the start of the next \task{} in the same execution thread, and simulate these gaps in Algorithm~\ref{algo:simulation}.

\textbf{Layer.} This field refers to which DNN layer a \task{} belongs to, which is necessary information for programmers to transform the graph and model optimizations. \system{} uses a synchronization-free approach to map a \task{} to DNN layers. We will describe the details of this approach in Section~\ref{sec:impl:map}.

\vspace{-2mm}

\subsubsection{Dependency} \label{sec:impl:dep}
\vspace{-1mm}

Based on our discussion in Section~\ref{sec:idea}, we identify the following five types of dependencies for accurate simulations.

\textbf{Sequential order of CPU \task{}s in the same thread.} CPU \task{}s in the same thread are serialized. The order that CPU \task{}s are executed in is determined by the framework and does not change in two separate executions. We add a dependency between each two consecutive CPU \task{}s in the same thread.

\textbf{Sequential order of GPU \task{}s in the same CUDA stream.} GPU kernels belonging to the same CUDA stream are executed sequentially. Similar to CPU \task{}s, the order of GPU \task{}s in the same stream does not change between executions. Hence, two consecutive GPU \task{}s in the same CUDA stream have a dependency between them.

\textbf{Correlation from CUDA APIs to GPU kernels.} Each GPU kernel or CUDA memory copy has a corresponding CPU-sided CUDA API (\texttt{cudaLaunch}, \texttt{cudaMemcpy}, or \texttt{cudaMemcpyAsync}) that triggers the GPU \task{}. CUPTI provides a correlation ID for every CUDA API and GPU kernel. A GPU kernel is dependant on a CUDA API if they share the same correlation ID.

\textbf{CUDA Synchronization.} A CUDA synchronization API (e.g., \texttt{cudaDeviceSynchronize}) is invoked on CPU, and returns after GPU kernels (or CUDA memory copies) that are launched before this synchronization complete. A CUDA synchronization therefore generates dependency from a GPU \task{} to a CPU \task{}. Similar to CUDA synchronizations, even though a \texttt{cudaMemcpyAsyncDtoH} call returns before a memory copy completes, we found it still blocks the CPU until all previous GPU kernels on the same stream are completed.

\textbf{Communication.} Mainstream frameworks including PyTorch and MXNet implement the wait-free backpropagation strategy~\cite{zhang2017poseidon} to schedule gradient communication. Here, a communication primitive is launched as soon as the weight gradients are ready, thus overlapping communication with the backward phases of subsequent layers. Hence, we need to know the runtime of DNN layers (not just kernels) to determine which tasks trigger communication.

\vspace{-2mm}
\subsection{Mapping Tasks to Layers} \label{sec:impl:map}
\vspace{-1mm}

\begin{figure}
    \centering
    \includegraphics[width=\columnwidth]{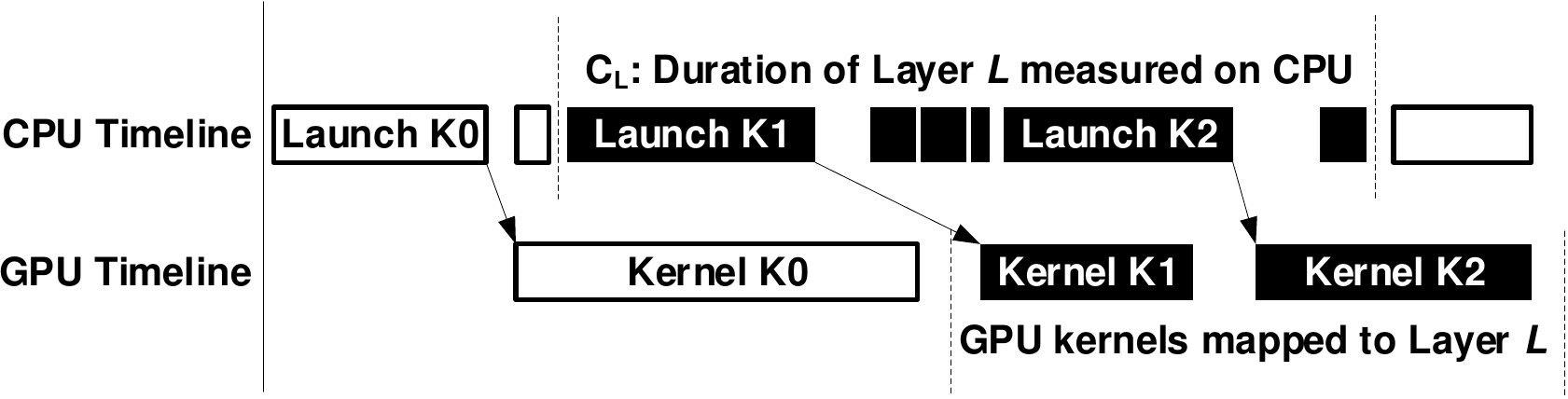}
    \caption{The mapping of GPU kernels to a layer. CUPTI provides correlations between CUDA launches and GPU kernels.}
    \label{fig:mapping}
    \vspace{-4mm}
\end{figure}

The task-to-layer mapping enables \system{} to construct the dependency graph for distributed training, and provides necessary domain knowledge for \system{} to model DNN optimizations.
Figure~\ref{fig:mapping} shows how \system{} determines which tasks belong to a certain layer. Let $L$ be the forward phase of a DNN layer. \system{} collects the CPU and GPU runtime information using CUPTI~\cite{cupti}, as well as timestamps before and after the forward, backward, and weight update phases for each layer. The start and end timestamps of $L$ will determine the CPU runtime of $L$ (denoted by $C_L$). To determine the GPU runtime of $L$, \system{} gathers all CUDA launch calls invoked during $C_L$. With CUPTI providing the correlations between CUDA launch calls and corresponding GPU kernels, \system{} can identify all the GPU kernels launched during $C_L$, and map these kernels to $L$.
This process can also be applied to the backward or weight update phases of any layers, and can be further generalized to any code region of interest in the framework or user-level programs. 

\vspace{-2mm}
\subsection{Graph Transformation}
\vspace{-1mm}
What-if analysis by transforming the graph and simulating the execution requires input about the optimizations from programmers. \system{} provides a set of primitives for programmers to model DNN optimizations by modifying the graph. Like most what-if analysis in non-ML contexts, modeling DNN optimizations requires potentially shrinking or scaling the duration of \task{}s (the \texttt{shrink/scale} primitives).
We carefully study common DNN optimization techniques and identify the following primitives (besides the \texttt{shrink/scale} primitives), which are sufficient for programmers to describe those optimizations.


\begin{figure}
    \centering
    \includegraphics[width=\columnwidth]{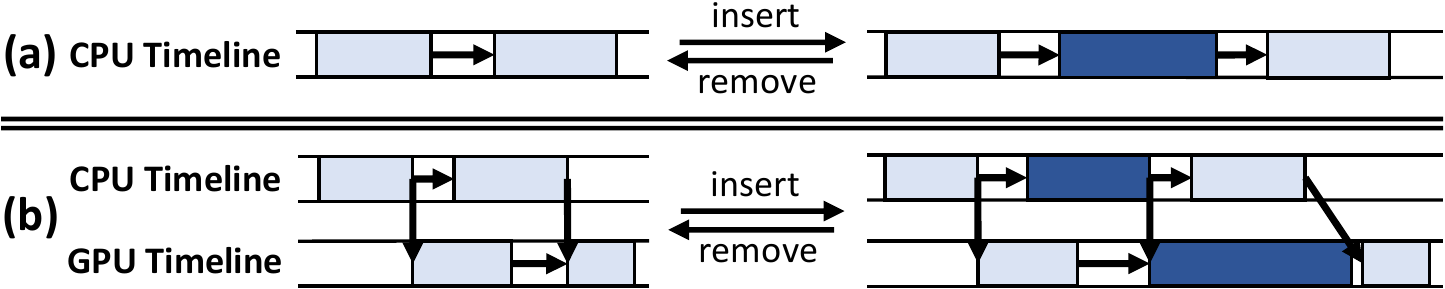}
    \caption{Insert/Remove a (a) CPU task; (b) GPU task.}
    \vspace{-4mm}
    \label{fig:insert_remove}
\end{figure}

\textbf{Insert/Remove a \task{}.} Inserting a \task{} to an execution thread just involves an appending of a node to a linked list. Figure~\ref{fig:insert_remove} shows how this process works. When inserting a GPU \task{}, we need to insert the corresponding CPU \task{}s that launch it. Which CPU \task{}s to insert and their duration depend on the framework implementation, and can be inferred based on collected traces.

\textbf{Select.} This operation allows users to select \task{}s of interest for further operations. One potentially useful selection criterion is select-by-layer, as many optimizations are depicted based on DNN layers. Another potentially useful criterion is to select by keywords in \task{} names, based on knowledge of the software library (e.g., cuDNN~\cite{cudnn}). For example, kernels with keywords such as \texttt{elementwise} or \texttt{PointwiseApply} in the names are element-wise arithmetic operations. These kernels are typically \emph{not} compute-bound, and could be much shorter than their corresponding CUDA launch calls. Similarly, kernels with \texttt{sgemm} string in names are compute-bound matrix-multiplications.

\textbf{Schedule.} The \texttt{schedule} function picks one \task{} from a set of frontier \task{}s that are ready to execute (line 9 in Algorithm~\ref{algo:simulation}). By default, it picks the task with the earliest start. Programmers can override this function and implement any custom scheduling policy. which is useful to model optimizations that increase computation-communication overlap.

%% file: sections/methodology.tex
\vspace{-2mm}
\section{Modeling Optimizations} \label{sec:reduce}
\vspace{-1mm}

To demonstrate that \system{} is able to estimate the performance of the most common optimizations in DNN training, we select ten techniques from Table~\ref{table:approches} with different optimization goals. We show that we can easily model these optimizations using the primitives \system{} provides
\footnote{We show pseudo code for AMP in this section.  Refer to the appendix for the pseudo code of all examples shown in Section~\ref{sec:reduce}.}.

\vspace{-2mm}
\subsection{Optimizations for Evaluation} \label{sec:reduce:eval}

We select the following five important optimizations to evaluate \system{}’s prediction accuracy.
We use implementations from the authors of these optimizations in cases where they were not readily available.

\textbf{Automatic Mixed Precision (AMP).} We aim to predict the efficacy of the AMP optimization~\cite{micikevicius2017mixed}, implemented using NVidia's Apex package~\cite{apex}. We expect that AMP will improve memory-bounded GPU kernels by 2$\times$ because the number of transferred bits is halved. With Tensor Cores in the Volta and Turing architectures, AMP empirically yields up to 3$\times$ speedup on the most compute-intensive workloads~\cite{amp-doc}.
To predict AMP performance, we simply \texttt{select} all the compute-intensive (e.g., \texttt{sgemm}, \texttt{conv}) kernels and memory-bounded (e.g., \texttt{elementwise}, \texttt{batchnorm}, \texttt{RELU}) kernels, and \texttt{shrink} their duration by 3$\times$ and 2$\times$ respectively.

\vspace{-2mm}
\IncMargin{1em}
\begin{algorithm}
\DontPrintSemicolon
\SetAlgoLined
\Indm
\SetKwInOut{Input}{Input}\SetKwInOut{Output}{Output}
\Input{Dependency graph: $G(V, E)$}
\Output{A modified graph $G(V, E)$ to model AMP}
\BlankLine
\Indp

$GPUTasks \gets \{G.Select(funcPtr(IsOnGPU))\}$

\ForEach{$u \in GPUTasks$} {
    \uIf {$"sgemm"$ in $u.Name$ or $"scudnn"$ in $u.Name$} {
        $u.duration \gets u.duration / 3$
    }
    \Else {
        $u.duration \gets u.duration / 2$
    }
}
\caption{What\_If\_AMP}\label{algo:what_if_amp}
\end{algorithm}
\vspace{-2mm}

\textbf{FusedAdam Optimizer.} We use the FusedAdam optimizer~\cite{apex_opt} implemented in NVidia's Apex package~\cite{apex} as an example for the kernel fusion optimization. This optimizer fuses all kernels in one weight update phase into one unified kernel. It is applicable to the models that use the Adam optimizer (e.g., GNMT, BERT). \system{} uses the kernel-to-layer mapping to identify the CPU/GPU tasks that belong to a weight update phase. We \texttt{remove} all these \task{}s, then \texttt{insert} a new GPU \task{} whose duration is roughly estimated by the sum of all removed compute-intensive kernels.

\textbf{Reconstructing Batchnorm.} Recently Jung et al.~\cite{jung2018restructuring} proposed a technique that optimizes non-convolutional layers in state-of-the-art CNNs. It first splits each batch normalization layer into two sub-layers, then fuses the first sub-layer with the previous convolutional layer, and the second sub-layer with the following activation and convolutional layers.
We \texttt{remove} the affected activation kernels when estimating performance, since they are memory-bound kernels now fused with compute-intensive convolutional kernels.
For the batch nomalization layers, we estimate that the GPU kernels will be improved by 2$\times$ since this optimization halves the amount of input data that these layers load from GPU memory.

\textbf{Distributed Training.} Using \system{} we can accurately predict distributed training performance with the profile based on the single-GPU environment. We evaluate \system{}'s prediction based on PyTorch, which uses collective communication primitives from the NCCl library~\cite{nccl}. PyTorch groups gradients from multiple layers into buckets before transferring them. Hence, to predict distributed training performance, we need to \texttt{insert} one allReduce \task{} for every bucket. The dependencies of the inserted \task{}s are determined based on the layer-to-bucket mapping (which requires additional instrumentation to the PyTorch framework).

\textbf{Priority-Based Parameter Propagation (P3).} P3~\cite{jayarajan2019priority} is a technique that optimizes communication overhead by slicing and prioritizing. We evaluate \system{}'s prediction of P3 based on MXNet, which uses the parameter-server mechanism~\cite{li2014scaling}. In order to model parameter slicing, we \texttt{insert} multiple push \task{} and pull \task{}s between the backward and the forward GPU \task{}s for each layer. The duration of the push/pull \task{} is calculated from the slice size and the network bandwidth. To model the priority scheduling, we override the \texttt{schedule} function with a priority queue.

\vspace{-2mm}
\subsection{Modeling Additional Optimizations} \label{sec:reduce:more}

In addition to the above optimizations, we show that \system{} is capable of modeling an additional set of diverse DNN optimizations.



\textbf{BlueConnect.} BlueConnect~\cite{cho2019blueconnect} optimizes communication by decomposing the allReduce primitives into a series of reduce-scatter and all-gather primitives. These primitives run concurrently as they use parallel communication channels.
To predict the performance of BlueConnect, instead of \texttt{insert}ing regular allReduce or push/pull \task{}s, we need to \texttt{insert} reduce-scatter and all-gather \task{}s, and assign them to corresponding network channels (the duration can be estimated according to formulas shown in \cite{nccltest}).

\textbf{MetaFlow.} MetaFlow~\cite{jia2019optimizing} is a layer-fusion technique to optimize DNN training by fusing DNN layers to simplify the DNN topology. 
We \texttt{select} the GPU kernels of substituted layers, \texttt{remove} them, and \texttt{insert} GPU kernels of new layers to predict the performance of MetaFlow in \system{}. The new layers are mostly existing layers with different dimensions; their GPU kernel durations can be inferred by profiling.

\textbf{vDNN.} Virtualized DNN~\cite{vdnn} reduces GPU memory consumption by temporarily offloading intermediate data from GPU memory to CPU memory. The offloaded data needs to be prefetched back to GPU to perform execution, which causes potential performance overhead due to PCIe traffic or late prefetching.
To predict the performance overhead using \system{}, we only need to \texttt{insert} additional CUDA memory copies, and override the \texttt{schedule} function to implement a custom prefetching policy.

\textbf{Gist.} Gist~\cite{gist} reduces GPU memory consumption by storing encoded intermediate data and decoding before the data is used. The encoding and decoding introduces performance overhead.
We \texttt{insert} extra encoding and decoding GPU kernels (along with \texttt{cudaLaunchKernel} calls in CPU) to estimate the performance overhead in \system{}. The duration of the inserted encoding/decoding kernels can be estimated using existing element-wise kernels.

\textbf{Deep Gradient Compression (DGC).} DGC~\cite{lin2017deep} is a technique that reduces communication overhead by compressing the gradients. To estimate performance, we: (i) \texttt{scale} the duration of communication; (ii) \texttt{insert} the GPU tasks of compression and decompression.
The duration of inserted GPU \task{}s can be estimated according to the compression rate and duration of existing element-wise GPU kernels.

%% file: sections/evaluation.tex
\vspace{-3mm}
\section{Evaluation} \label{sec:eval}


\vspace{-2mm}
\subsection{Methodology}
\vspace{-1mm}

We implement \system{} based on three mainstream DNN frameworks: PyTorch~\cite{pytorch}, MXNet~\cite{mxnet}, and Caffe~\cite{jia2014caffe}. We add CUPTI~\cite{cupti} support to each framework to obtain traces of CUDA APIs and GPU kernels. We also add instrumentation to the frameworks to acquire layer-wise timestamps for the kernel-to-layer mapping process, and communication information such as the size of each allReduce call and their dependencies with other layer-wise computation.

\textbf{Infrastructure.} We evaluate \system{}'s runtime prediction on a cluster of four machines. Each machine contains one AMD EPYC 7601 16-core processor~\cite{amd-epyc}, and four 2080Ti GPUs~\cite{2080ti} with 11GB GDDR6 memory each, connected through PCIe 3.0~\cite{ajanovic2008pci}. Our experiments are based on Ubuntu 16.04, CUDA v10.0~\cite{cuda-10}, cuDNN v7.4.1~\cite{cudnn-7_4_1}, and NCCL v2.4.2~\cite{nccl}. Our software implementation is based on PyTorch v1.0, MXNet v1.1, and Caffe v1.0.

\begin{table}
\ra{1.1}
\centering
\small
\begin{tabular}{lll} 
\toprule
Application                           & Model     & Dataset                   \\ \hline
Image Classification                  & VGG19~\cite{simonyan2014very}     & ImageNet~\cite{deng2009imagenet}                  \\
                                      & DenseNet-121~\cite{huang2017densely}  &                           \\
                                      & ResNet-50~\cite{resnet} &                           \\ \hline
Machine Translation                   & GNMT~\cite{gnmt}   & WMT16~\cite{wmt16}                     \\ \hline
Language Modeling                     & BERT~\cite{devlin2018bert}      & SQuAD~\cite{rajpurkar2016squad}                     \\
\bottomrule
\end{tabular}
\vspace{-2mm}
\caption{The models and datasets we use in this paper.}
\label{table:bench}
\end{table}

\textbf{Models.} Table~\ref{table:bench} shows the DNN models and datasets we use to evaluate \system{}. We select five DNN models from three different applications, covering a diverse set of DNN models. For the BERT model, we evaluate both "base" and "large" versions. The difference between these versions is that the "base" version contains 12 "Transformer blocks" (the main layer type in BERT) where as the "large" version contains 24.

\vspace{-2mm}
\subsection{Automatic Mixed Precision (AMP)} \label{sec:eval:amp}
\vspace{-1mm}

\begin{figure}
    \centering
    \includegraphics[width=0.85\columnwidth]{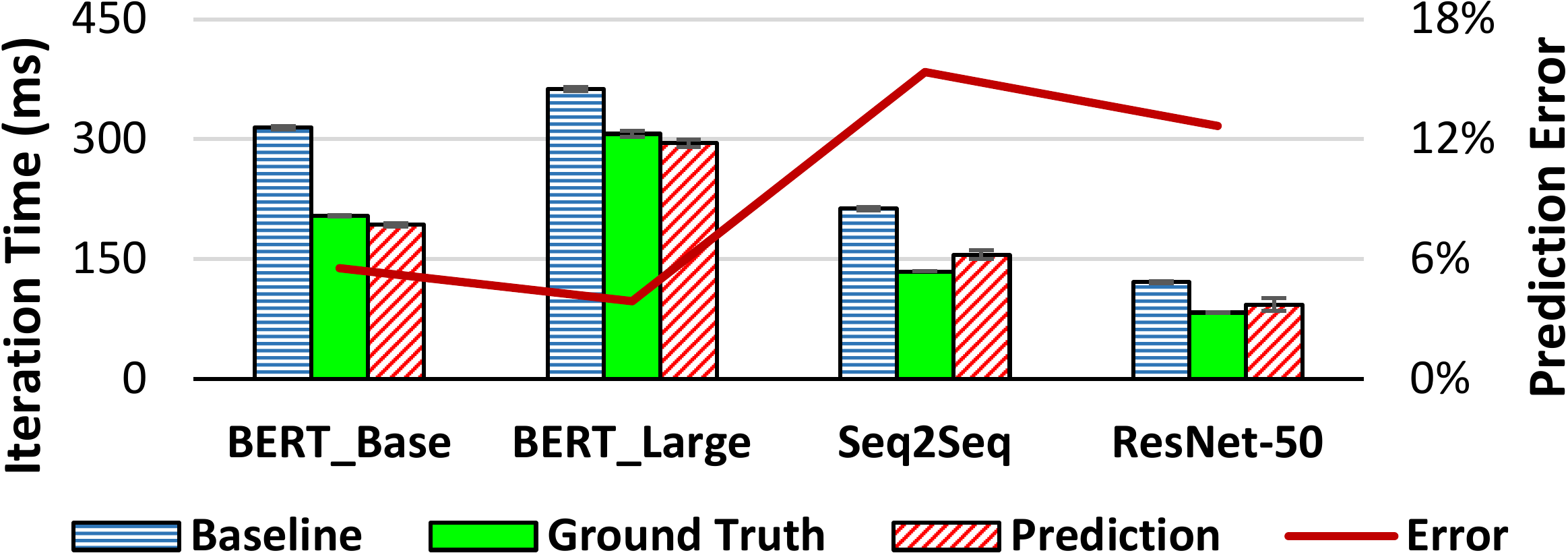}
    \caption{AMP \--- comparing baseline (FP32), ground truth with mixed precision, and predictions by \system{}.}
    \vspace{-3mm}
    \label{fig:mixed_precision}
\end{figure}

We evaluate \system{}'s prediction accuracy of AMP~\cite{micikevicius2017mixed}, which is implemented in NVidia's Apex package~\cite{apex} based on the PyTorch framework. Figure~\ref{fig:mixed_precision} shows the performance of using AMP and the corresponding performance prediction given by \system{}. Our predictions have errors below 13\% for all the models we evaluate.

\begin{figure}
    \centering
    \includegraphics[width=0.85\columnwidth]{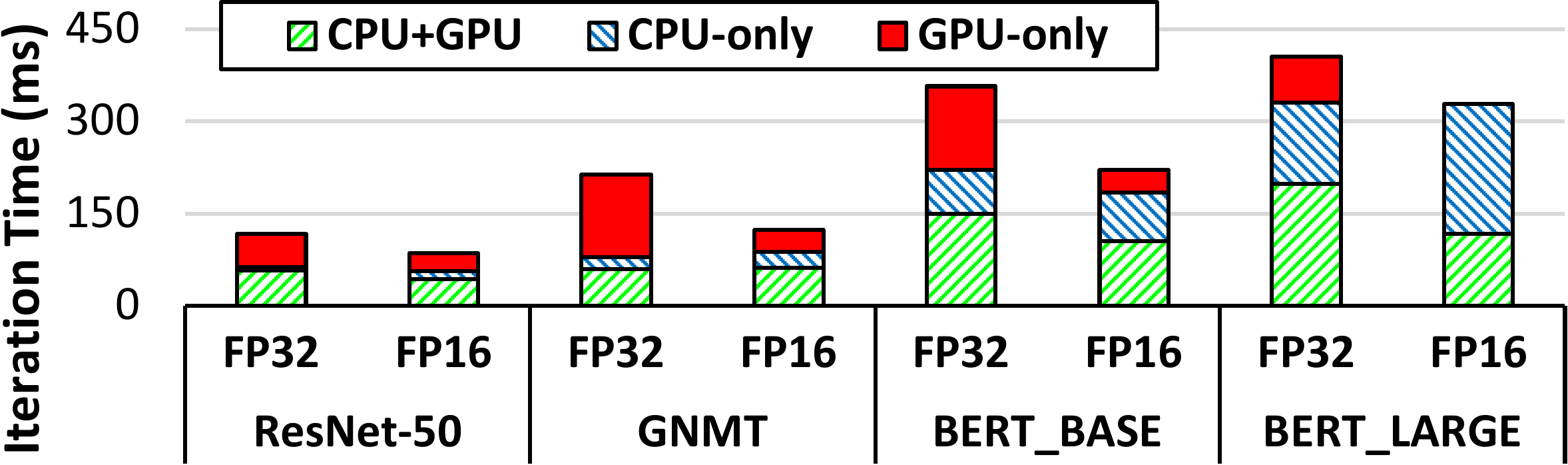}
    \caption{Runtime breakdown of the baseline (FP32) and mixed precision (FP16).}
    \label{fig:mixed_precision_breakdown}
    \vspace{-2mm}
\end{figure}

Our experiments show that using AMP brings speedups generally less than 2$\times$ \--- much less than the theoretical boost of using AMP for individual kernels (e.g., 3$\times$). To understand how AMP improves  performance, we break down the overall runtime into the following three components:

\textbf{CPU-only runtime.} This component refers to the runtime when the CPU is busy, but the GPU is not executing any kernels. It is straightforward to calculate this runtime by simply subtracting all GPU kernel runtime from the total runtime.

\textbf{GPU-only runtime.} This component refers to the runtime when the CPU is waiting for the GPU kernels to complete. It includes not only the duration of CUDA synchronization APIs, but also the \texttt{cudaMemcpyAsync} calls of all the device-to-host CUDA memory copies.

\textbf{CPU+GPU parallel runtime.} This component refers to the runtime when both CPU and GPU are busy. We calculate this part of runtime by deducting the CPU-only and GPU-only parts from the total runtime.

Figure~\ref{fig:mixed_precision_breakdown} shows the runtime breakdown of the models we evaluated. CPU runtime generally becomes the new performance bottleneck in the models that incur limited speedups (e.g., BERT\textsubscript{LARGE}). When applying AMP, the CPU bottleneck increases, because the GPU runtime becomes shorter and part of the CPU+GPU parallel runtime is shifted to the CPU-only runtime. The overall runtime improvement comes mostly from the reduction of GPU-only runtime while CPU runtime barely changes. This demonstrates the necessity of the kernel-level abstraction when predicting performance.

\vspace{-2mm}
\subsection{FusedAdam Optimizer}
\vspace{-1mm}

\begin{figure}
    \centering
    \includegraphics[width=0.85\columnwidth]{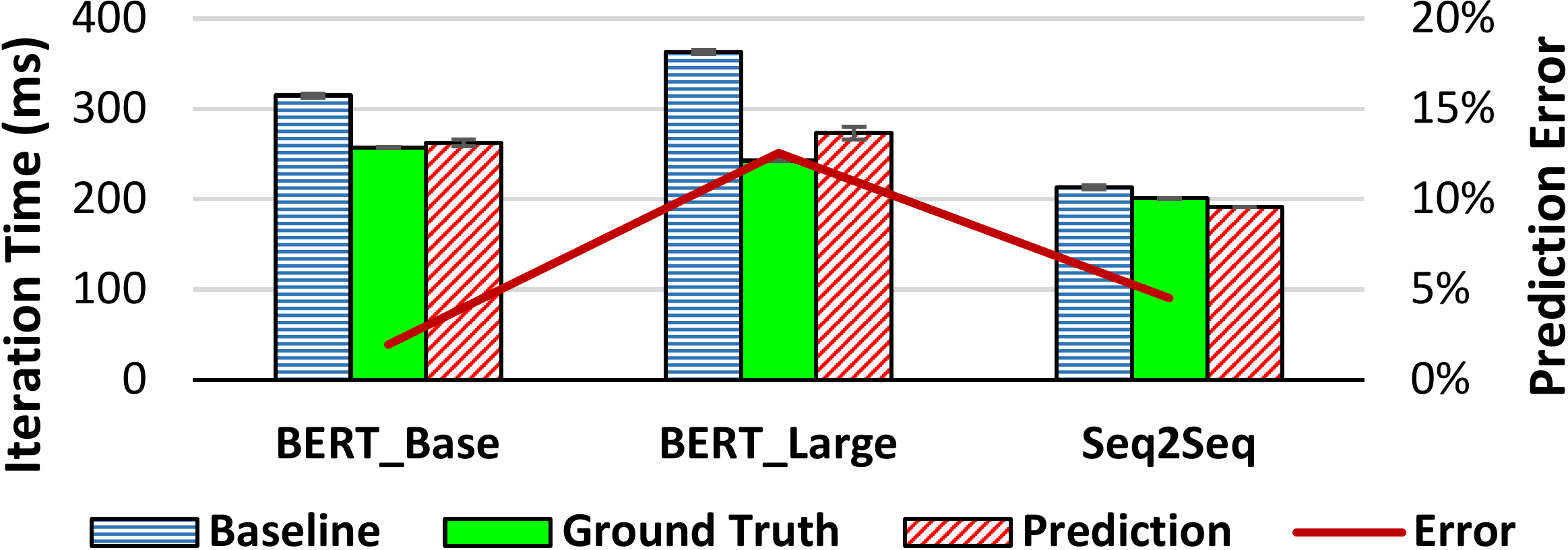}
    \caption{FusedAdam - comparing baseline (FP32), ground truth with FusedAdam, and predictions by \system{}.}
    \vspace{-3mm}
    \label{fig:fused_adam}
\end{figure}

We apply the FusedAdam optimization to the BERT and GNMT models as they use the Adam optimizer. Figure~\ref{fig:fused_adam} shows the performance of using the FusedAdam optimizer. Our predictions are within 13\% of the ground truth runtime.



There are two reasons why the FusedAdam optimizer substantially improves the performance of BERT models.
First, unlike most DNN training workloads, the weight update phase is a significant proportion of a BERT model's iteration runtime (around 30\% for BERT\textsubscript{BASE} and 45\% for BERT\textsubscript{LARGE}).
Second, the weight update phase consists of very many element-wise GPU kernels (2633 for BERT\textsubscript{BASE}, 5164 for BERT\textsubscript{LARGE}). Thus, the CUDA launch calls on the CPU become the main bottleneck.
The FusedAdam optimizer almost eliminates all CPU kernel launch overhead in the weight update phase by fusing all GPU kernels into one single GPU kernel. Compared to BERT models, the GNMT model spends less than 10\% of its iteration time on the weight update phase, explaining the lower speedup improvements.

\begin{figure}[t]
    \centering
    \subfloat[Runtime predictions for ResNet-50.\label{fig:sync_reduce_resnet}]{\includegraphics[width=\columnwidth]{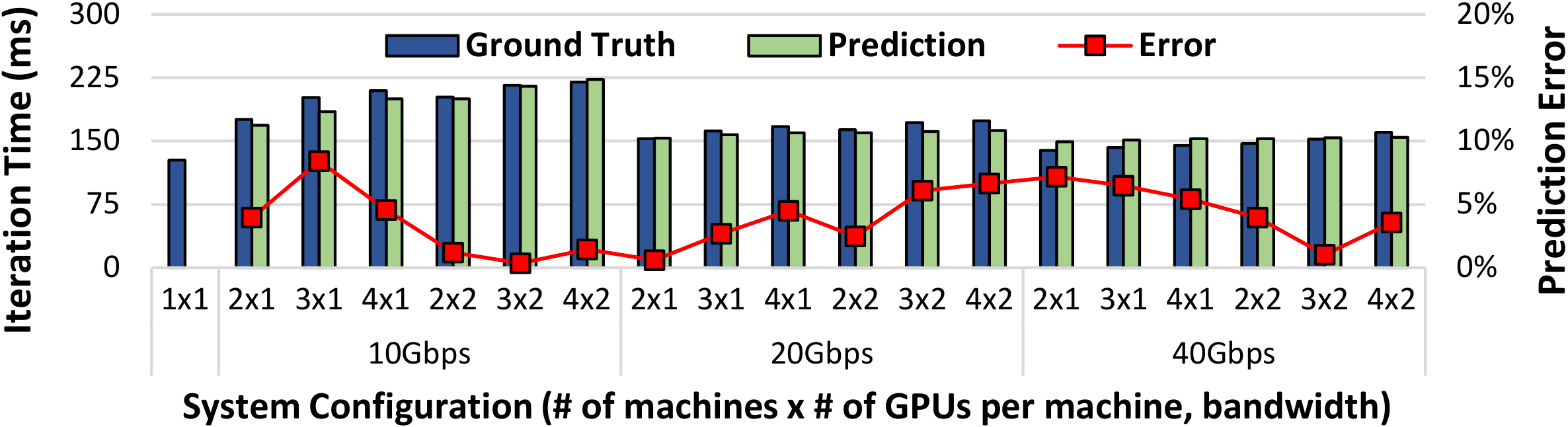}}
    \vspace{-1mm}
    
    \subfloat[Runtime predictions for GNMT.\label{fig:sync_reduce_gnmt}]{\includegraphics[width=\columnwidth]{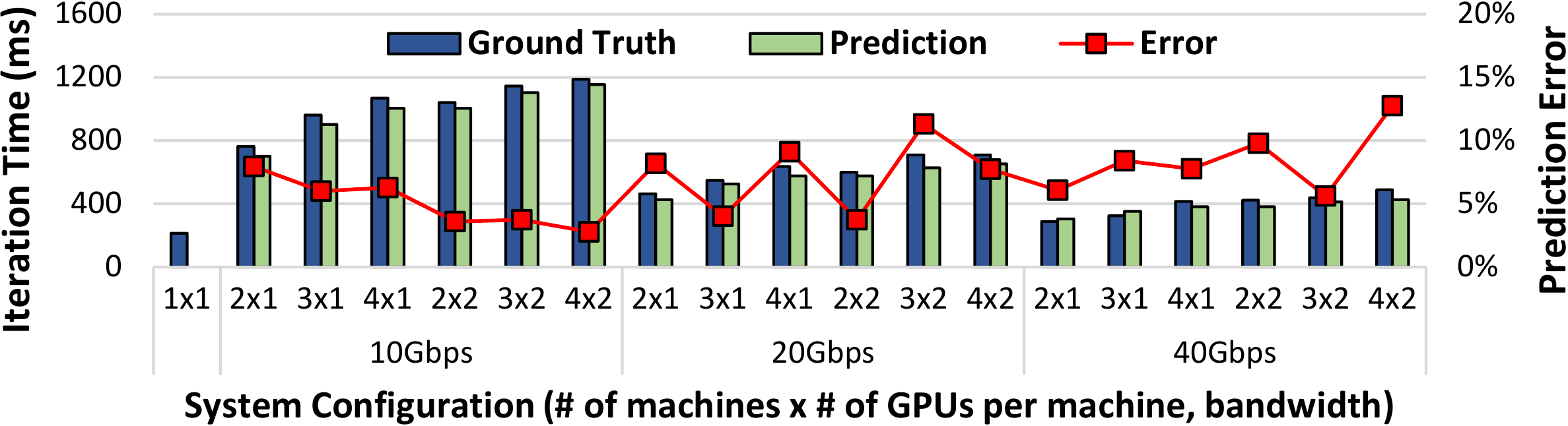}}
    \vspace{-1mm}
    
    \subfloat[Runtime predictions for BERT\textsubscript{BASE}.\label{fig:sync_reduce_bert_base}]{\includegraphics[width=\columnwidth]{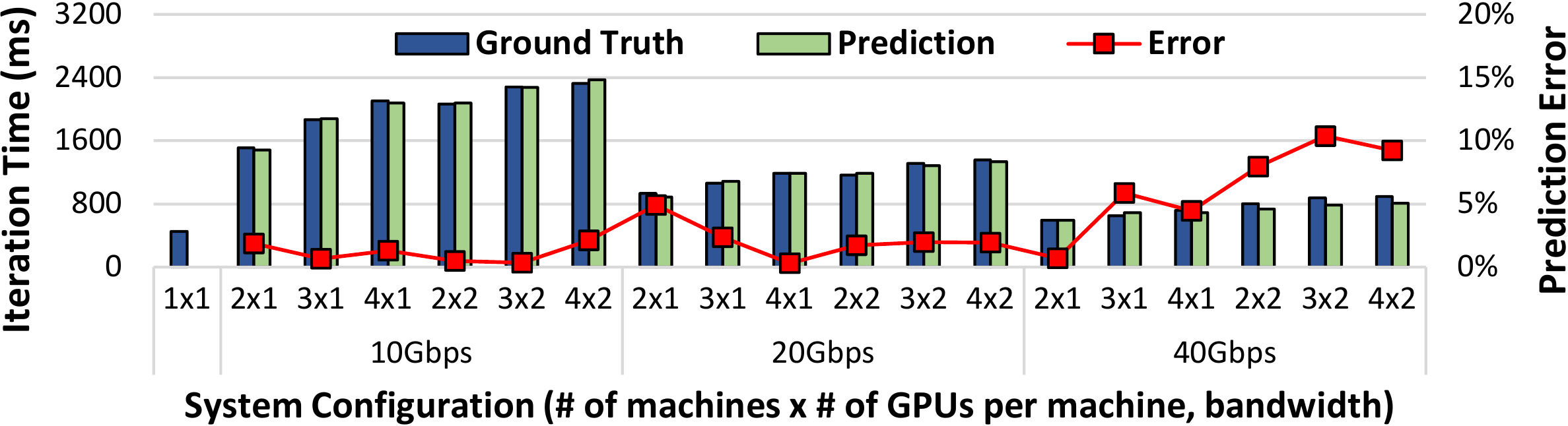}}
    \vspace{-1mm}
    
    \subfloat[Runtime predictions for BERT\textsubscript{LARGE}.\label{fig:sync_reduce_bert_large}]{\includegraphics[width=\columnwidth]{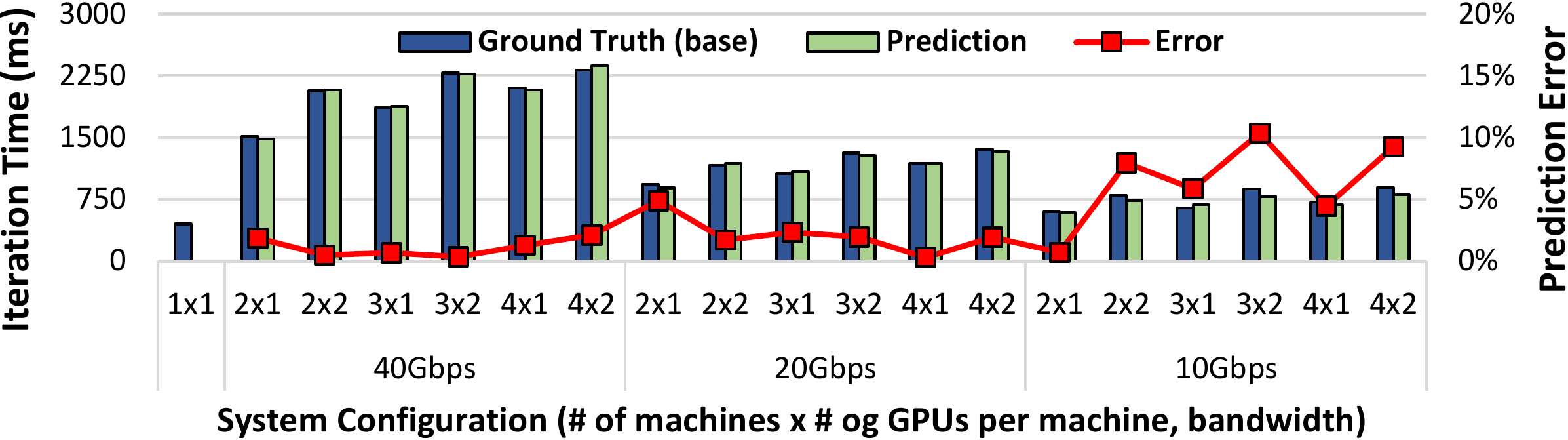}}
    \vspace{-1mm}
    
    \caption{The error between \system{}'s runtime predictions and the baseline with synchronization before each allReduce under various system configurations.}
    \vspace{-4mm}
    \label{fig:predict_scalalibity}
\end{figure}

\vspace{-2mm}
\subsection{Reconstructing Batchnorm} \label{sec:eval:batchnorm}
\vspace{-1mm}

We evaluate our performance prediction for the optimization of reconstructing batch normalization~\cite{jung2018restructuring} based on the Caffe implementation of DenseNet-121~\cite{huang2017densely}. Using \system{}, we predict that reconstructing batchnorm will yield a moderate performance improvement of 12.7\% compared to the baseline. This suggests that reconstructing batchnorm in our configuration is less promising than the paper claims (17.5\% speedup). We verify this conclusion by testing the ground truth implementation of reconstructing batchnorm, and find out that this optimization yields even lower 7\% speedup.

We notice that there are two main reasons for the difference between our prediction and the ground truth. First, the ground truth uses a completely new implementation of the batchnorm layers, and it is hard to precisely predict the runtime of newly implemented kernels. Second, the ground truth implementation introduces new CUDA memory copies and allocations, which add performance overhead. Obtaining a very precise estimate would require us to understand not just the high-level idea from the paper, but also the detailed implementation of the user-level programs and the Caffe framework.

\vspace{-2mm}
\subsection{Distributed Training}
\vspace{-1mm}

Next we evaluate distributed training using PyTorch with the NCCL~\cite{nccl} library. Figure~\ref{fig:predict_scalalibity} shows the comparisons between runtimes predicted by \system{} and the measured ground truth runtimes, for each DNN model under different system configurations. We evaluate the prediction accuracy for Ethernet and InfiniBand connecting multi-machine systems under different network bandwidths (10, 20, 40 Gbps). In most of the configurations, \system{} predicts distributed runtime with at most 10\% prediction error, with a few exceptions for the 20Gbps and 40Gbps configurations.

\begin{figure}
    \centering
    \includegraphics[width=\columnwidth]{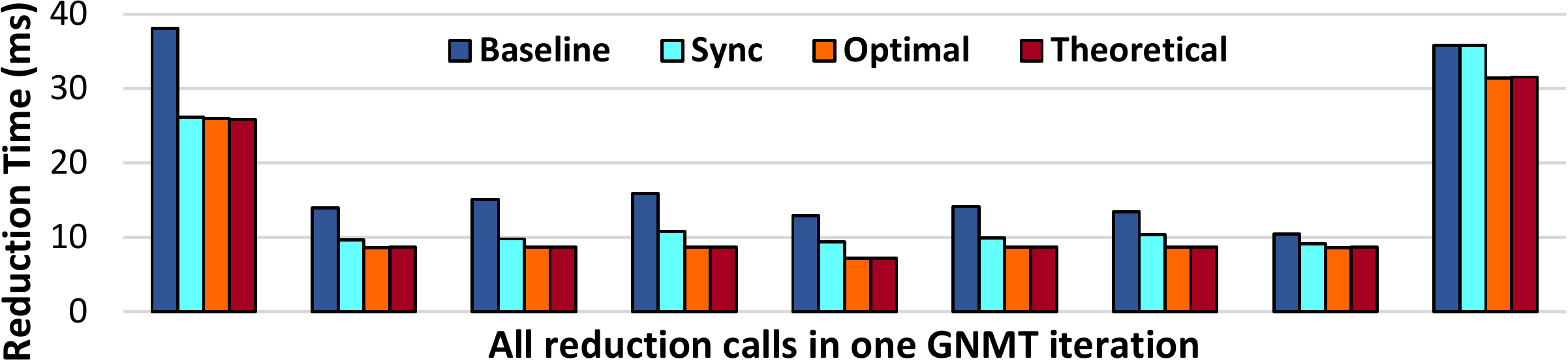}
    \caption{Comparison of all individual reduction runtimes in one training iteration of GNMT. \textbf{Baseline}: runtime measured in regular training; \textbf{Sync}: runtime measured with an additional CUDA synchronization before each reduction; \textbf{Optimal}: runtime measured when executing exclusively; \textbf{Theoretical}: runtime calculated using the formula~\cite{nccltest}.}
    \vspace{-4mm}
    \label{fig:allreduce}
\end{figure}

The prediction errors of the overall iteration times are mainly due to inaccurate estimates of individual NCCL primitives. Figure~\ref{fig:allreduce} shows the comparisons of NCCL allReduce calls between the ground truths and predictions. The ground truths are on average 34\% higher than the theoretical values.

An NCCL primitive is both a communication primitive and a GPU kernel, suggesting that it could be bottlenecked by two types of hardware resources: (i) the network bandwidth, and (ii) GPU resources (e.g., memory bandwidth, streaming multiprocessors). 
Figure~\ref{fig:allreduce} shows that the predicted values are very close to the runtimes measured when running NCCL primitives exclusively. This suggests that the ground truth is slower because they compete for GPU resources with other GPU kernels.
Based on this insight, we try to reduce this interference by adding CUDA synchronizations before invoking NCCL primitives.
As shown in Figure~\ref{fig:allreduce}, adding synchronizations improve the NCCL primitives by 22.8\% on average when compared to the baseline.

We also verify the impact to the overall iteration time when adding synchronizations before NCCL primitives. We run the experiments on all the configurations shown in Figure~\ref{fig:predict_scalalibity}. We find that this simple approach does not lead to performance degradation in any configuration. Instead, it could bring an improvement of up to 22\%.

\vspace{-2mm}
\subsection{Priority-Based Parameter Propagation}
\vspace{-1mm}

\begin{figure}
    \centering
    \subfloat[ResNet-50.\label{fig:p3_resnet}]{\includegraphics[width=0.48\columnwidth]{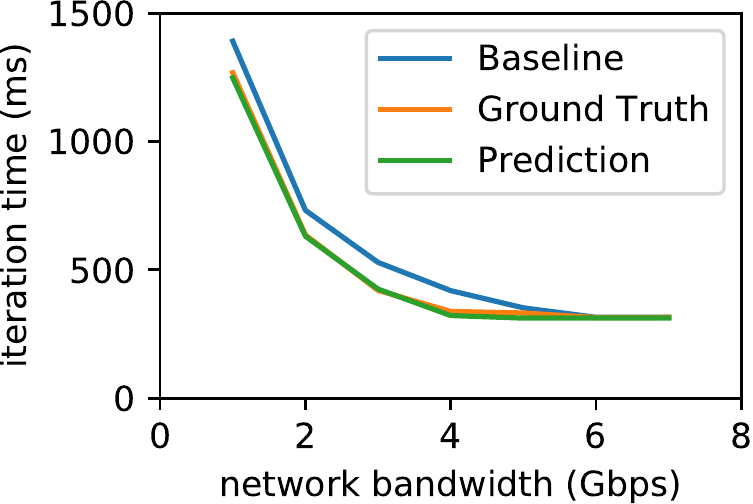}}
    \subfloat[VGG-19.\label{fig:p3_vgg}]{\includegraphics[width=0.48\columnwidth]{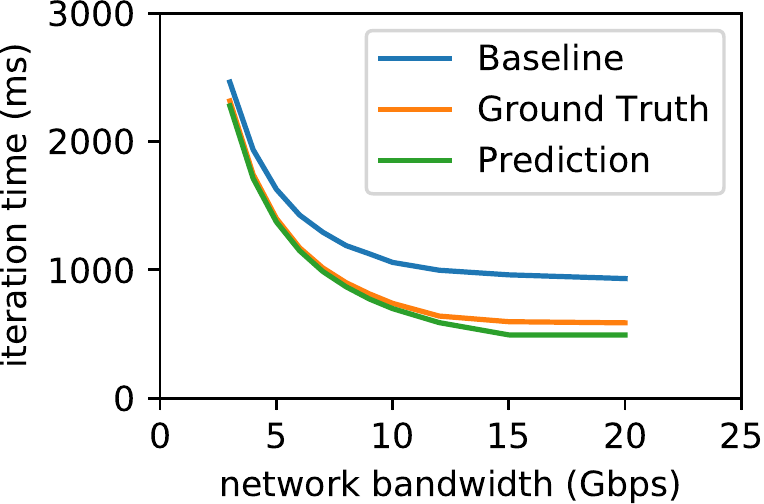}}
\vspace{-3mm}
    \caption{\system{}'s prediction for how the P3 optimization will help under different network bandwidths.}
    \vspace{-5mm}
    \label{fig:p3_pred}
\end{figure}

We evaluate \system{}'s prediction accuracy of applying Priority-Based Parameter Propagation (P3) to VGG-19 and ResNet-50. To reproduce the performance speedups of P3, we use a cluster of four machines with one P4000 GPU per machine (which is consistent with the evaluation setup of the P3 paper~\cite{jayarajan2019priority}). We use MXNet v1.1, and have one worker process and one parameter server process on each machine.

Figure~\ref{fig:p3_pred} shows the iteration time of the baseline, ground truth, and prediction using \system{} under different bandwidths. Our prediction faithfully reflects the trend of P3 speedups when the network bandwidth increases. The prediction error is at most 16.2\% among all the configurations we tested, and lower in most of the configurations.

We overestimate the speedup of P3, especially when training VGG-19 with a 15 or 20 Gbps network bandwidth. The reason is similar to our previous insight about NCCL primitives: when bandwidth is higher, a communication task is increasingly bottlenecked by non-network resources. In the case of MXNet, this overhead could be caused by the server processes, or the control flow of the worker processes.

%% file: sections/discussion.tex
\vspace{-2mm}
\section{Discussion}
\vspace{-2mm}

In this section, we discuss the adaptability, potential extensions, and some limitations of Daydream.

\vspace{-2mm}
\subsection{Why Not Simply Run the Optimizations?}
\vspace{-1mm}

The main problem many ML developers face is that not all optimizations are readily available on all platforms. In fact, we are only able to evaluate the prediction accuracy of optimizations with the implementations already available (see Table~\ref{table:approches}); for the remaining ones, we highlight the flexibility of Daydream by showing that they can be represented succinctly. Most newly proposed optimizations do not have open-source implementations on \textbf{all} DNN frameworks available right away; it would be unreasonable to expect researchers to open-source their implementations and port their optimizations on all platforms. Therefore, analyzing if these optimizations can help in a deployment setting, using Daydream, can still precede the programming effort to port the optimizations.  Furthermore, Daydream's profiling can be performed just once, and using that profile on a given platform, one can answer questions for many different optimizations.

\vspace{-2mm}
\subsection{Adaptability of Daydream}
\vspace{-1mm}

Daydream requires support from hardware profilers. The current implementation of Daydream utilizes GPU-based profilers, and it relies on CUPTI to provide: (i) CPU and GPU traces and (ii) information about which CPU call triggered the launch of a specific GPU kernel. Adapting our design to other architectures (e.g., TPUs), would require hardware vendor profilers to provide similar traces for this new hardware.

Daydream can be also easily adapted to other ML frameworks (e.g., MXNet and TensorFlow). We built Daydream based on PyTorch, and then post-process the dumped traces to make predictions. The post-processing scripts are framework-independent. To add framework instrumentation, we need to: (i) add CUPTI (or similar tool) support, (ii) insert per-layer timestamps, and (iii) gather the gradient-to-bucket mappings for injecting the communication primitives to the dependency graph (required for PyTorch). Such instrumentation is relatively light-weight and can be easily adapted to other mainstream frameworks such as TensorFlow~\cite{tf} and MXNet~\cite{mxnet}.



\vspace{-2mm}
\subsection{Training Accuracy Prediction}
\vspace{-1mm}
In addition to improving iteration time, some optimizations may also affect training accuracy (e.g., AMP~\cite{micikevicius2017mixed}, DGC~\cite{lin2017deep}); predicting the impact of optimizations on accuracy is currently outside of Daydream's scope. We leave this interesting and challenging problem for future work.

\vspace{-2mm}
\subsection{Kernel Runtime Prediction}
\vspace{-1mm}
Estimating the effect of optimizations that alter existing GPU kernels or introduce new ones requires predicting the runtime of new/changed GPU kernels. When estimating performance of AMP, our estimation of kernels that use half-precision kernels was based on findings/observations from NVIDIA~\cite{micikevicius2017mixed}. This generalization above for all kernels (in contrast to identifying how each kernel in isolation is affected by AMP), still leads to the low prediction errors we observe in Figure~\ref{fig:mixed_precision}. 

However, optimizations such as DGC~\cite{lin2017deep}, Reconstructing Batchnorm~\cite{jung2018restructuring}, and Gist~\cite{gist} introduce newly-implemented kernels to the runtime.
Accurately predicting runtime for new kernels is a challenging problem. 
Daydream estimates the overall runtime based on existing kernel implementations, or using guidelines from studies that highlight quantitative improvements for the proposed kernels.  
But if the estimated runtimes for such new kernels are inaccurate, it may lead to relatively high prediction error (Section~\ref{sec:eval:batchnorm}). How much a kernel's runtime estimation error contributes to the overall prediction error depends on the training workload itself.

While Daydream cannot predict individual kernel runtime, it provides a high-level structure for kernel developers to estimate the overall performance. Developers can profile their individual kernels, and then input the profiling results into Daydream to accurately estimate the overall runtime. This approach saves the engineering effort of porting the kernel implementation into the DNN frameworks.

\vspace{-2mm}
\subsection{Concurrent Kernels}
\vspace{-1mm}

Existing GPU profilers such as CUPTI usually serialize GPU kernel execution, removing all concurrency, making our performance estimation somewhat conservative. Despite this, we observe that the runtime for models with concurrent execution (e.g., GNMT) can still be predicted with high accuracy (\S~\ref{sec:eval:amp}). This is because the majority of computation time goes to fully connected layers (including embedding layers), which have no concurrent kernels executed in parallel with them. 
We leave a complete solution for concurrent kernels, requiring better support from profiling tools, as a part of future work.

%% file: sections/related.tex
\vspace{-2mm}
\section{Related Work}
\vspace{-2mm}

To help programmers understand the performance of the hardware accelerators and develop highly efficient applications, hardware vendors provide profiling tools (e.g., NVProf~\cite{nvprof}, Nsight~\cite{nsight}, vTune~\cite{reinders2005vtune}) that can reveal low-level performance counters (e.g., cache hit rate, memory speed, clock rate). These tools are usually designed with general applications in mind, and expose hundreds of low-level performance counters. The fundamental limitation of all these tools is that they do not utilize application-specific knowledge.


The new generation of profiling tools feature the \emph{application-aware} property, enabling them to deliver domain-specific (e.g., ML-specific) insights about performance to programmers. The Cloud TPU Tool~\cite{tputool} is an example of such a profiling tool. It correlates low-level TPU metrics with the DNN structure, and shows the performance for each DNN layer.
Similarly, MXNet~\cite{mxnet} and PyTorch~\cite{pytorch} also have their own built-in profiling tools. These domain-specific tools can highlight performance hotspots, but are less efficient in finding optimization opportunities. 
In contrast, \system{} is not only \emph{application-aware}, but also \emph{optimization-aware}, enabling \system{} to quantitatively estimate the efficacy of different optimizations without fully implementing them.

Prior works have tried to explore what-if questions in other contexts by using low-level traces.
Curtsinger~\emph{et al.} proposed a causal profiler (COZ~\cite{curtsinger2015c}) to identify potentially unknown optimization opportunities by running performance simulation with certain functions being virtually speed-up. Unlike \system{}, COZ does not require dependencies among functions because it does not consider the cases where functions can be added or deleted (which is the case for many ML optimizations). Pourghassemi~\emph{et al.} uses the idea of COZ to analyze the performance for web browser applications~\cite{pourghassemi2019if}. 
For data analytic frameworks, such as Spark~\cite{zaharia2010spark}, Ousterhout~\emph{et al.} use dependency analysis to understand the overhead caused by I/O, network, and stragglers~\cite{ousterhout2015making, ousterhout2017monotasks}. 
\system{} is designed to address a more diversified set of what-if questions, and hence requires more powerful modeling.

Prior works address what-if questions of the form "What if we can speedup task $T$ by $N$ times (or infinity)?", but they do not study whether existing optimizations can deliver this speedup.
In the ML context, given an optimization, accurately predicting the performance of individual tasks in the dependency graph, is still an open problem.
It requires additional knowledge about the kernel implementation and the architecture design. Currently \system{} can not automatically estimate the runtime of new GPU kernels.
However, as we show in Section~\ref{sec:eval}, even with rough estimates of per-kernel duration based on domain knowledge and reasonable assumptions, we can still achieve high overall prediction accuracy.


%% file: sections/conclusion.tex
\vspace{-3mm}
\section{Conclusion}
\vspace{-2mm}

The efficacy of DNN optimizations can vary largely across different DNN models and deployments. \system{} is a new profiler to effectively explore the efficacy of a diverse set of DNN optimizations.
\system{} achieves this goal by using three key ideas: (i) constructing a kernel-level dependency graph by utilizing vendor-provided profiling tools, while tracking dependencies among concurrently executing tasks; (ii) mapping low-level traces to DNN layers in a synchronization-free manner; (iii) introducing a set of rules for programmers to effectively describe and model different optimizations.
Our evaluation shows that using \system{}, we can effectively model (i.e. predict runtime) the most common DNN optimizations, and accurately identify both optimizations that result in significant performance improvements as well as those that provide limited benefits or even slowdowns.

%% file: sections/acknowledgement.tex
\vspace{-2mm}
\section*{Acknowledgement}
\vspace{-1mm}

Daydream is part of Project Fiddle at MSR. We thank the MSR
Lab LT, especially Ricardo Bianchini and Donald Kossmann, for
their enthusiastic and unwavering support of Project Fiddle.
We also thank our shepherd, Swaminathan Sundararaman, the anonymous ATC reviewers, Brian Hirano, James Gleeson, Geoffrey Yu, Xiaodan (Serina) Tan, Jorgen Thelin, Shivaram Venkataraman, and Deepak Narayanan, for their constructive feedback during the development of this work. This work was also supported in part by the NSERC Discovery grant, the Canada Foundation for Innovation JELF grant, the Connaught Fund, and Huawei grants.

%% file: sections/appendix.tex
\appendices

\section{Modeling Optimizations}

Due to the space limitation, we are not able to include all details in our main sections. The most important ones are about how to use them to model various optimizations. In this appendix we provide these details.

As shown in Table~\ref{table:approches}, there are a wide range of DNN optimizations, which would introduce various impacts on the training runtime. One of such impacts is that duration of tasks in will scale/shrink. For example, using AMP will shrink the duration of GPU kernels. Using Daydream, such impact is easy to model with the help of the \texttt{Select} operator to pick tasks of interests.

DNN optimizations might alter the network topology (e.g. kernel fusion~\cite{ashari2015optimizing}, MetaFlow~\cite{jia2019optimizing}), TASO~\cite{jia2019taso}, introduce new operators (e.g. Gist~\cite{gist}, vDNN~\cite{vdnn}, Deep Gradient Compression~\cite{lin2017deep}), or restructuring the communication scheme (e.g., P3~\cite{jayarajan2019priority}, BlueConnect~\cite{cho2019blueconnect}). These optimizations will eventually alter the low-level dependency graph, adding or removing GPU kernels and communication primitives. Daydream provides \texttt{Insert/Remove} operators for programmers to model these transformations. Programmers need to locate where tasks are inserted/removed with the help of the \texttt{Select} operator. As we will show later, this locating varies across different optimizations, but is generally not complicated.

Rescheduling tasks is another transformation that needs to be supported in Daydream. This operator does not change the dependency graph topology or the task duration. Instead, it manipulates the execution order of the tasks, and aims at higher parallelism among the tasks. One example of such transformation is the prioritization scheme in P3~\cite{jayarajan2019priority}. Modeling this scheme involves just overriding the \texttt{Scheduling} function in the simulation process~\ref{algo:simulation}. Programmers might need to attach additional attributes to the tasks to implement a custom scheduling policy. In the optimizations we show below, modeling P3~\cite{jayarajan2019priority} and vDNN~\cite{vdnn} require overriding the \texttt{Scheduling} function.

\subsection{Automatic Mixed Precision (AMP)}

To model AMP, we shrink the duration of GPU kernels by 2$\times$. If TensorCore is available on the GPU, compute intensive kernels such as \texttt{sgemm} are expected to speed up by 3$\times$~\cite{amp-doc}. We show the pseudo code in Algorithm~\ref{algo:what_if_amp}.

\vspace{-2mm}
\IncMargin{1em}
\begin{algorithm}
\DontPrintSemicolon
\SetAlgoLined
\Indm
\SetKwInOut{Input}{Input}\SetKwInOut{Output}{Output}
\Input{Dependency graph: $G(V, E)$}
\Output{An updated graph $G(V, E)$ to model AMP}
\BlankLine
\Indp

$GPUTasks \gets \{G.Select(funcPtr(IsOnGPU))\}$

\ForEach{$u \in GPUTasks$} {
    \uIf {$"sgemm"$ in $u.Name$ or $"scudnn"$ in $u.Name$} {
        $u.duration \gets u.duration / 3$
    }
    \Else {
        $u.duration \gets u.duration / 2$
    }
}
\caption{What\_If\_AMP}\label{algo:what_if_amp}
\end{algorithm}
\vspace{-2mm}

\subsection{Fused Adam Optimizer}

The Fused Adam optimizer fuses all the kernels in the weight update phase. To model this optimizer, we remove all but one kernels in the weight update phase, and scale the duration of the remaining kernels with the sum of all fused ones. We show the pseudo code in Algorithm~\ref{algo:what_if_fused_adam}.

\vspace{-2mm}
\begin{algorithm}
\DontPrintSemicolon
\SetAlgoLined
\Indm
\SetKwInOut{Input}{Input}\SetKwInOut{Output}{Output}
\Input{Dependency graph: $G(V, E)$}
\Output{Am updated graph $G(V, E)$ to model the Fused\_Adam optimizer}
\BlankLine
\Indp

$GPUTasks \gets \{G.Select(funcPtr(IsOnGPU))\}$

$WUTasks \gets {GPUTasks.Select(funcPtr(IsWeightUpdate))}$ \\

$WUSum \gets 0$ \\

\ForEach{$u \in WUTasks$} {
    $WUSum \gets WUSum + u.duration$
}

$First \gets True$ \\

\ForEach{$u \in WUTasks$} {
    \uIf {First}
    {
        $u.duration \gets WUSum$ \\
        $First \gets False$
    }
    \uElse{
        $G.Remove(u)$
    }
}
\caption{What\_If\_Fused\_Adam}\label{algo:what_if_fused_adam}
\end{algorithm}

\subsection{Reconstructing Batchnorm}

Reconstructing Batchnorm~\cite{jung2018restructuring} improves the performance of training CNNs by splitting batch normalization layers and fusing memory-intensive kernels with compute-intensive kernels. We show the pseudo-code of using Daydream to model this optimization. We show the pseudo code in Algorithm~\ref{algo:what_if_batchnorm}.

\vspace{-2mm}
\begin{algorithm}
\DontPrintSemicolon
\SetAlgoLined
\Indm
\SetKwInOut{Input}{Input}\SetKwInOut{Output}{Output}
\Input{Dependency graph: $G(V, E)$}
\Output{An updated graph $G(V, E)$ to model Restructuring\_Batchnorm}
\BlankLine
\Indp

$GPUTasks \gets \{G.Select(funcPtr(IsOnGPU))\}$

\ForEach{$u \in GPUTasks$}{
    \If{$u.layer$ is ReLU}{
        $G.Remove(u)$
    }
    \If{$u.layer$ is Batchnorm} {
        $u.duration \gets u.duration / 2$
    }
}

\caption{What\_If\_Restructuring\_Batchnorm}\label{algo:what_if_batchnorm}
\end{algorithm}
\vspace{-2mm}

\subsection{Distributed Training}

We show how to use Daydream to model distributed training in PyTorch's decentralized architecture with the NCCL backend, based on runtime on a single GPU. When invoking NCCL all-reduce primitives, PyTorch groups small gradient tensors together to better utilize the bandwidth. Such grouping information can be collected by instrumentation from the PyTorch framework. In our code example, we use \texttt{layer\_bucket\_id} to represent the mapping from layers to communication buckets. Each bucket corresponds to one communication call. We show the pseudo code in Algorithm~\ref{algo:what_if_distributed}.

\vspace{-2mm}
\begin{algorithm}
\DontPrintSemicolon
\SetAlgoLined
\Indm
\SetKwInOut{Input}{Input}\SetKwInOut{Output}{Output}
\Input{Dependency graph: $G(V, E)$, Gradient Grouping: layer\_bucket\_id[]}
\Output{An updated graph $G(V, E)$ to model distributed training}
\BlankLine
\Indp

$GPUTasks \gets \{G.Select(funcPtr(IsOnGPU))\}$

$Bucket\_Task \gets []$ \\

$WU \gets$ the earliest node in the weight update phase \\

\ForEach{$b \in [1..\#\_of\_bucket]$}{
    $AllReduceTask = newNode("AllReduce", ...)$ \\
    $AllReduceTask.size \gets 0$ \\
    $G.AddDependencies(AllReduceTask \rightarrow{} WU)$ \\
    $Bucket\_Task[n] \gets AllReduceTask$
}

\ForEach{$u \in GPUTasks$}{
    \If{$u$ is FF\_layer}{
        $bucket\_id \gets layer\_bucket\_i[u]$ \\
        $T \gets Bucket\_Task[bucket\_id]$ \\
        $T.size \gets T.size + u.gradient\_size$ \\
        $G.AddDependencies(u \rightarrow{} T)$
    }
}

\caption{What\_If\_Distributed\_Training}\label{algo:what_if_distributed}
\end{algorithm}

\subsection{Priority-based Parameter Propagation (P3)}

P3~\cite{jayarajan2019priority} splits each gradient tensor into small slices and reschedules the communication based on the order in which gradient tensors are generated. We show how to model P3 based on MXNet's parameter server architecture (with push/pull communication primitives). To model P3 with Daydream, we insert parallel push/pull primitives for each gradient slice, tag each slice with priority based on the generation order, and override the \texttt{Schedule} function to model the prioritization scheme.

\vspace{-2mm}
\begin{algorithm}
\DontPrintSemicolon
\SetAlgoLined

\Indm
\SetKwInOut{Input}{Input}\SetKwInOut{Output}{Output}
\Input{Dependency graph: $G(V, E)$, slice\_size}
\Output{An updated graph $G(V, E)$ to model P3}
\BlankLine
\Indp

\tcp{Select GPU \task{}s in BP and FF}

$GPUTasks \gets \{G.Select(funcPtr(IsOnGPU))\}$

\ForEach{$u \in GPUTasks$} {
    $v \gets u$'s corresponding BP layer \\
    
    $g \gets$ |$u.layer$'s gradients|

    \While {$g > 0$} {
        $s \gets$ min($g$, $slice\_size$)

        $push \gets$ newNode("push v.layer", s, ...)
        
        $pull \gets$ newNode("pull v.layer", s, ...)
        
        $push.priority \gets$ -(distance to output)

        $push.ExecutionThread \gets comm.send$

        \uIf {this slice is stored on the first server} {
            $pull.ExecutionThread \gets comm.send$
        }
        \uElse{
            $pull.ExecutionThread \gets comm.receive$
        }
            
        $G.AddDependencies(u \rightarrow{} push \rightarrow{} pull \rightarrow{} v)$
            
        $g \gets g - slice\_size$
    }
}

\SetKwFunction{FMain}{Schedule}
\SetKwProg{Fn}{Function}{:}{}

\Fn{\FMain{TaskQueue: $Q$}}{
    $earliest \gets Q.first()$ \\
    $thread \gets earliest.ExecutionThread$ \\
    $time \gets max(P[thread], earliest.start)$ \\
    
    \ForEach{$task \in Q$}{
        $this\_thread \gets task.ExecutionThread$ \\
        $this\_time \gets max(P[this\_thread], task.start)$ \\
        \If{$this\_time < time$}{
            $time \gets this\_time$ \\
            $earliest \gets task$
        }
        \If{$this\_time = time \land task$ is $push/pull \land earliest$ is $push/pull \land task.priority > earliest.priority$}{
            $earliest \gets task$
        }
    }

    \Return $earliest$
}

\textbf{End Function}

\caption{What\_If\_P3}\label{algo:what_if_p3}

\end{algorithm}
\DecMargin{1em}
\vspace{-2mm}

\subsection{BlueConnect}

BlueConnect~\cite{cho2019blueconnect} optimizes the bandwidth usage by decomposing the synchronous all-reduce operations into a series of reduce-scatter and all-reduce operations. The decomposition helps better utilize the heterogeneous intra-node and inter-node bandwidths. The decomposition of all-reduce operations is based on a factorization of the number of GPUs. We show the pseudo code in Algorithm~\ref{algo:what_if_blueconnect}.

\vspace{-2mm}
\begin{algorithm}
\DontPrintSemicolon
\SetAlgoLined
\Indm
\SetKwInOut{Input}{Input}\SetKwInOut{Output}{Output}
\Input{Dependency graph of distributed training: $G(V, E)$, decomposition factorization: $p_1p_2...p_k$}
\Output{Am updated graph $G(V, E)$ to model BlueConnect}
\BlankLine
\Indp

$ReduceTasks \gets \{G.Select(funcPtr(IsAllReduce))\}$

\ForEach{$u \in ReduceTasks$} {
    $s \gets u.prevNodes$ \\
    $t \gets u.postNodes$ \\
    $G.Remove(u)$ \\
    \ForEach{$i \gets 1..k$}{
        $RSNode \gets new(Reduce\_Scatter\_Node(p_i))$ \\
        $G.Insert(s, RSNode, t)$ \\
        $s \gets RSNode$ \\
    }
    \ForEach{$i \gets k..1$}{
        $AGNode \gets new(All\_Gather\_Node(p_i))$ \\
        $G.Insert(s, AGNode, t)$ \\
        $s \gets AGNode$ \\
    }
}

\caption{What\_If\_BlueConnect}\label{algo:what_if_blueconnect}
\end{algorithm}
\vspace{-2mm}

\subsection{MetaFlow}

MetaFlow~\cite{jia2019optimizing} is a relaxed graph substitution optimizer. It simplifies the layer representation of a DNN topology by using operations like enlarging convolution kernel dimensions and layer fusion. The policy to transform the layer-wise topology is determined by a backtracking search algorithm. Daydream does not provide extra support that automatically determines the policy, as this is a duplicated work.

A transformation policy of MetaFlow will eventually remove or scale the dimension of existing layers. Given a policy, Daydream can estimate its performance by modeling layer-wise removal/scaling operations, with the help of layer mapping (described in Section~\ref{sec:impl:map}). We show Daydream's pseudo code of implementing these two operations in Algorithm~\ref{algo:what_if_metaflow}.

\begin{algorithm}
\DontPrintSemicolon
\SetAlgoLined
\Indm
\SetKwInOut{Input}{Input}\SetKwInOut{Output}{Output}
\Input{Dependency graph: $G(V, E)$}
\Output{An updated graph $G(V, E)$ to model MetaFlow}
\BlankLine
\Indp

\SetKwFunction{RMain}{Remove\_layer}
\SetKwProg{Fn}{Function}{:}{}

\Fn{\RMain{Dependency Graph: G(V, E), Layer: l}}{
    $GPUTasks \gets \{G.Select(funcPtr(IsOnGPU))\}$ \\
    \ForEach{$u \in GPUTasks$}{
        \If{$u.layer$ is $l$}{
            $G.Remove(u)$
        }
    }
}
\textbf{End Function}

\SetKwFunction{SMain}{Scale\_layer}
\SetKwProg{Fn}{Function}{:}{}

\Fn{\SMain{Dependency Graph: G(V, E), Layer: l}}{
    $GPUTasks \gets \{G.Select(funcPtr(IsOnGPU))\}$ \\
    \ForEach{$u \in GPUTasks$}{
        \If{$u.layer$ is $l$}{
            $u.duration \gets u.duration \times s$
        }
    }
}
\textbf{End Function}

\caption{What\_If\_MetaFlow}\label{algo:what_if_metaflow}
\end{algorithm}

MetaFlow's search algorithm uses a cost model to evaluate the performance of a given policy. Daydream can be used as a more precise cost model for the search algorithm.

\subsection{Virtualized DNN (vDNN)}

Virtualized DNN~\cite{vdnn} optimizes the memory footprint in CNN training by offloading feature maps from GPU memory to CPU memory. To model vDNN with Daydream, we only need to insert the corresponding cudaMemcpy calls, and implement prefetching strategy by using the overriding \texttt{Schedule} function. The custom \texttt{Schedule} function delays the execution of the prefetching operation. We demonstrate how to model the \texttt{vDNN\textsubscript{\textit{conv}}} policy, which only offloads the feature maps of all convolutional layers. We tag each layer with an ID (a layer with higher ID means closer to the output layer), and use the \texttt{findPrefetchLayer} function defined in the original vDNN paper~\cite{vdnn}. We show the pseudo code in Algorithm~\ref{algo:what_if_vdnn}.

\vspace{-2mm}
\begin{algorithm}
\DontPrintSemicolon
\SetAlgoLined
\Indm
\SetKwInOut{Input}{Input}\SetKwInOut{Output}{Output}
\Input{Dependency graph: $G(V, E)$}
\Output{An updated graph $G(V, E)$ to model vDNN}
\BlankLine
\Indp

$GPUTasks \gets \{G.Select(funcPtr(IsOnGPU))\}$ \\

$ID2PrefetchTask \gets \{\}$ \\

\ForEach{$u \in GPUTasks$}{
    \If{$u.layer$ is not CONV\_FF}{
        $continue$
    }
    $v \gets u$'s corresponding BP layer \\
    $t1 \gets newCPUNode("cudaMemcpyLaunch", ...)$ \\
    $t2 \gets newGPUNode("cudaMemcpyH2D", ...)$ \\
    $t3 \gets newCPUNode("cudaFree\_vDNN", ...)$ \\
    $t4 \gets newCPUNode("cudaMalloc\_vDNN", ...)$ \\
    $ID2PrefetchTask[u.ID] \gets t4$ \\
    $t5 \gets newCPUNode("cudaMemcpyLaunch", ...)$ \\
    $t6 \gets newGPUNode("cudaMemcpyD2H", ...)$ \\
    $G.addDependencies(u \rightarrow{} t1 \rightarrow{} t2 \rightarrow{} t3 \rightarrow{} t4 \rightarrow{} t5 \rightarrow{} t6 \rightarrow{} v)$
}

\SetKwFunction{FMain}{Schedule}
\SetKwProg{Fn}{Function}{:}{}

\Fn{\FMain{TaskQueue: $Q$}}{
    $GPUTasks \gets \{G.Select(funcPtr(IsOnGPU))\}$ \\
    
    $next \gets Q.last()$ \\
    
    \If{$next.layer$ is BP}{
        $l \gets findPrefetchLayer(next.ID)$ \\
        \uIf{$l \neq -1$}{
            \Return $next$
        }
        \Else{
            \Return $ID2PrefetchTask[l]$
        }
    }
}
\textbf{End Function}

\caption{What\_If\_vDNN}\label{algo:what_if_vdnn}
\end{algorithm}
\vspace{-2mm}

\subsection{Gist}

Gist~\cite{gist} is an technique that optimizes the memory footprint when training CNNs. It reduces the memory consumption of the intermediate feature maps by adding encoding/decoding operations to the training iterations. Gist provides both lossless and lossy compression strategies. We can use Daydream to estimate the performance overhead of Gist, by inserting the encoding/decoding kernels. When estimating the lossless compression, we need to insert GPU kernels that are either element-wise kernels (including clamping, pooling-mapping, bit-wise kernels, etc.), or cuSPARSE kernels. When estimating the lossy compression, we need to additionally insert the GPU kernels that perform Delayed Precision Reduction (DPR) scheme.

Note that estimating the duration of these kernels is crucial to the prediction accuracy. The duration of these kernels can be either inferred based on existing kernels, or profiled separately (the latter is outside of Daydream's focus and should be resolved using other techniques). We show the pseudo code in Algorithm~\ref{algo:what_if_gist}.

\vspace{-2mm}
\begin{algorithm}
\DontPrintSemicolon
\SetAlgoLined
\Indm
\SetKwInOut{Input}{Input}\SetKwInOut{Output}{Output}
\Input{Dependency graph: $G(V, E)$}
\Output{An updated graph $G(V, E)$ to model Gist}
\BlankLine
\Indp

$GPUTasks \gets \{G.Select(funcPtr(IsOnGPU))\}$ \\

\ForEach{$u \in GPUTasks$}{
    $v \gets u.postNode$ \\
    $w \gets v.postNode$ \\
    \If{$u.layer$ is RELU\_FF $\land v.layer$ is POOL\_FF$ \land w.layer$ is CONV\_FF}{
        $SSDC\_kernels \gets newNode(...)$ \\
        $G.insert(v, SSDC, w)$
    }
    \If{$u.layer$ is RELU\_FF $\land v.layer$ is POOL\_FF}{
        $Binarize \gets newNode(...)$ \\
        $G.insert(v, Binarize, w)$
    }
}

\If{LOSSY\_COMPRESSION}{
    \ForEach{$u \in GPUTasks$}{
        \If{$u$ is not RELU}{
            $v \gets u.postNode$ \\
            $DPR \gets newNode(...)$ \\
            $G.insert(u, DPR, v)$
        }
    }
}

\tcp{Add decode kernels to the backward pass} 

...

\caption{What\_If\_Gist}\label{algo:what_if_gist}
\end{algorithm}
\vspace{-2mm}

\vfill

\subsection{Deep Gradient Compression (DGC)}

DGC~\cite{lin2017deep} reduces communication overhead by compressing the gradients before transmission and decompressing the gradients before weight update phase. When using Daydream to estimate the performance overhead of DGC, we need to insert the compression/decompression kernels before/after the communication primitives. Similar to Gist, the prediction accuracy mainly depends on the estimation of the inserted kernels. We show the pseudo code in Algorithm~\ref{algo:what_if_dgc}.

\vspace{-2mm}
\begin{algorithm}
\DontPrintSemicolon
\SetAlgoLined
\Indm
\SetKwInOut{Input}{Input}\SetKwInOut{Output}{Output}
\Input{Dependency graph: $G(V, E)$}
\Output{An updated graph $G(V, E)$ to model Deep Gradient Compression}
\BlankLine
\Indp

$ReduceTasks \gets \{G.Select(funcPtr(IsAllReduce))\}$ \\

\ForEach{$r \in ReduceTasks$}{
    $s \gets r.prevNodes()$ \\
    $t \gets r.postNodes()$ \\
    \tcp{Initialize compression kernels}
    $quantize\_op \gets newNode(...)$ \\
    $sparse\_op \gets newNode(...)$ \\
    ... \\
    $G.Insert(s, quantize\_op, r)$ \\
    $G.Insert(quantize\_op, sparse\_op, r)$ \\
    ... \\

    \tcp{Initialize decompression kernels}
    $d\_kernels \gets ...$ \\
    $G.Insert(r, d\_kernels, t)$
}

\caption{What\_If\_DGC}\label{algo:what_if_dgc}
\end{algorithm}

\vfill\clearpage